%% file: main.tex
\documentclass[a4paper,11pt]{article}

\usepackage{jinstpub}
\usepackage{graphicx}
\usepackage{hyperref}
\usepackage{siunitx}
\usepackage{upgreek}
\usepackage[titletoc]{appendix}
\usepackage{multirow}
\usepackage{booktabs}
\usepackage{amsmath}
\usepackage[normalem]{ulem}
\usepackage[super]{nth}
\usepackage{comment}
\usepackage{xcolor}
\usepackage{lineno}

\newcommand{\pin}{$\textit{p-i-n }$}
\newcommand{\fref}[1]{figure~\ref{#1}}

\newcommand{\tref}[1]{table~\ref{#1}}

\newcommand{\sref}[1]{section~\ref{#1}}
\newcommand{\secref}[1]{Section~\ref{#1}}
\newcommand{\eref}[1]{equation~(\ref{#1})}

\title{Forward Beam Monitor for the KATRIN experiment}
\input{author-list.tex}

\abstract{The \textit{KArlsruhe TRItium Neutrino} (KATRIN) experiment aims to measure the neutrino mass with a sensitivity of \SI{0.2}{\electronvolt} (\SI{90}{\percent} CL). This will be achieved by a precision measurement of the endpoint region of the $\upbeta$-electron spectrum of tritium decay. The $\upbeta$-electrons are produced in the \textit{Windowless Gaseous Tritium Source} (WGTS) and guided magnetically through the beamline. In order to accurately extract the neutrino mass the source activity is required to be stable and known to a high precision. The WGTS therefore undergoes constant extensive monitoring from several measurement systems. The \textit{Forward Beam Monitor} (FBM) is one such monitoring system.

The FBM system comprises a complex mechanical setup capable of inserting a detector board into the KATRIN beamline with a positioning precision of better than \SI{0.3}{\milli\metre}. The electron flux density at that position is on the order of \SI{e6}{\per\second\per\milli\metre\squared}. The detector board contains two silicon detector chips of \pin diode type which can measure the $\upbeta$-electron flux from the source with a precision of \SI{0.1}{\percent} within \SI{60}{\second} with an energy resolution of FWHM $=$ \SI{2}{\kilo\electronvolt}.

The unique challenge in developing the FBM arise from its designated operating environment inside the Cryogenic Pumping Section which is a potentially tritium contaminated ultra-high vacuum chamber at cryogenic temperatures in the presence of a \SI{1}{\tesla} strong magnetic field. Each of theses parameters do strongly limit the choice of possible materials which e.g. caused difficulties in detector noise reduction, heat dissipation and lubrication. In order to completely remove the FBM from the beam tube a \SI{2}{\meter} long traveling distance into the beamline is needed demanding a robust as well as highly precise moving mechanism.}


\begin{document}

    \newpage
	\maketitle
	
	\section{Introduction}
	\label{Section:Introduction}
	
	The KATRIN experiment will improve the sensitivity of neutrino mass measurements to $m_{\nu} =$ \SI{0.2}{\electronvolt} (\SI{90}{\percent} C.L.) corresponding to a \SI{5}{\sigma} discovery potential for a mass signal of $m_{\nu} =$ \SI{0.35}{\electronvolt}~\cite{Osipowicz:2001,Angrik:2005} in the most sensitive direct neutrino mass experiment to date. The neutrino mass will be derived from a precise measurement of the shape of the tritium $\upbeta$-decay spectrum near its endpoint at $E_{0} =$ \SI{18573.7\pm0.1}{\electronvolt}~\cite{PRL2019}. The source of $\upbeta$-electrons is a \textit{Windowless Gaseous Tritium Source} (WGTS) which has an activity of \SI{e11}{\becquerel}.
	
\begin{figure}[!ht]
        \centering
        \includegraphics[width=1.0\textwidth]{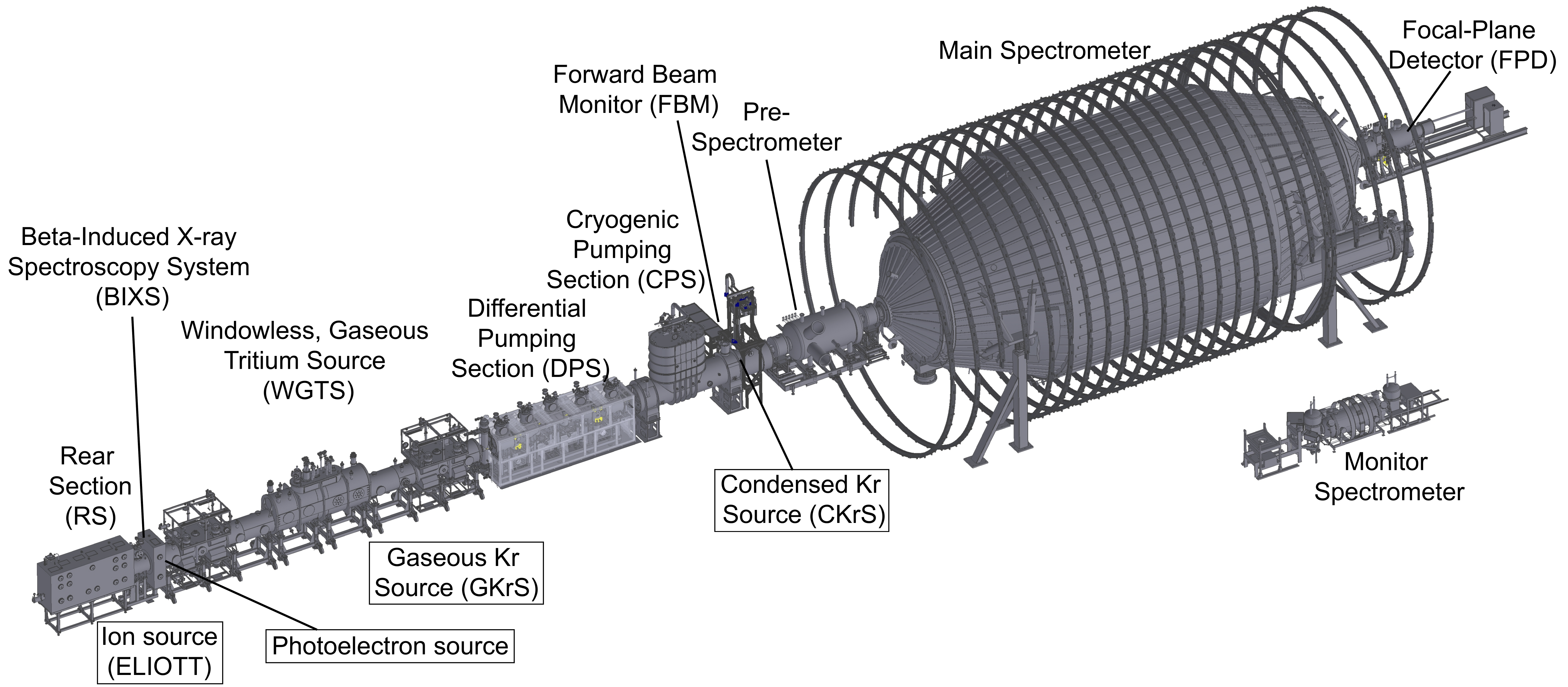}
        \caption{The KATRIN beamline. The FBM is located at the end of the CPS and represents the final source monitoring system before the $\upbeta$-electrons enter the spectrometer and detector section.}
        \label{Figure:KatrinBeamline}
	\end{figure}

	The layout of the KATRIN beamline \cite{Arenz_2016} is shown in \fref{Figure:KatrinBeamline}. The \textit{Source and Transport Section} (STS) consists of the WGTS, the \textit{Differential Pumping Section} (DPS), the \textit{Cryogenic Pumping Section} (CPS), and several source monitoring and calibration systems~\cite{Babutzka:2012}. Along the beamline superconducting solenoids generate a magnetic field of several Tesla strength which adiabatically guides the $\upbeta$-electrons towards the spectrometers while excess tritium is pumped out of the system. The \textit{Spectrometer and Detector Section} (SDS) consists of the pre-spectrometer, the main-spectrometer, the monitor-spectrometer, and the \textit{Focal Plane Detector} (FPD). All spectrometers are of MAC-E-Filter type which transmit electrons with energies above a chosen retarding energy \cite{PICARD1992345}, and reject those with lower energies. The main-spectrometer can perform an energy analysis of the $\upbeta$-electrons with an energy resolution of \SI{0.93}{\electronvolt} at \SI{18.6} {\kilo\electronvolt}.

	 The source-related parameters associated with the main systematic uncertainties in the determination of the neutrino mass are activity fluctuations of the WGTS, energy loss corrections (of $\upbeta$-electron scattering in the WGTS), the final state distribution, the source magnetic field, and the source plasma condition.
	
	In order to analyse the tritium $\upbeta$-spectrum and determine the neutrino mass the WGTS needs to be extremely stable, particularly in its activity and  isotopic composition. Therefore, the WGTS properties need to be known with high precision, and are continuously monitored for short and long term fluctuations. There are several monitoring and calibration subsystems associated with the WGTS~\cite{Babutzka:2012}.
	
	Results from the various subsystems are combined over long time periods during extended measurement time. This paper focuses on one such activity monitoring system, the \textit{Forward Beam Monitor} (FBM). The FBM is the final monitoring subsystem for $\upbeta$-electrons from the source before they enter into the spectrometer and detector section. It has been commissioned prior to the KATRIN krypton measurement campaign in June 2017 \cite{Arenz_2018}. Initial data was then obtained during the krypton measurement campaign and during the KATRIN first tritium measurement campaign in May 2018 \cite{Firsttritium}. The FBM is capable of continuously monitoring variations of the electron flux and changes in the measured shape of the $\upbeta$-decay spectrum during the KATRIN neutrino mass measurement phases.

	This paper is organised as follows. In \sref{Section:TritiumSource} the WGTS and its operating parameters are introduced and in \sref{Section:MeasurementPrinciple} the FBM measurement principle for the monitoring of the relevant WGTS parameters is explained. \secref{Section:TechnicalDescription} contains a technical description of the FBM. In \sref{Section:Measurements} the FBM commissioning and results from the krypton and first tritium measurement phases are presented, and \sref{Section:Conclusion} contains the summary.

	\section{Tritium source}
	\label{Section:TritiumSource}
	
\begin{figure}[!ht]
        \centering
        \includegraphics[width=0.8\textwidth]{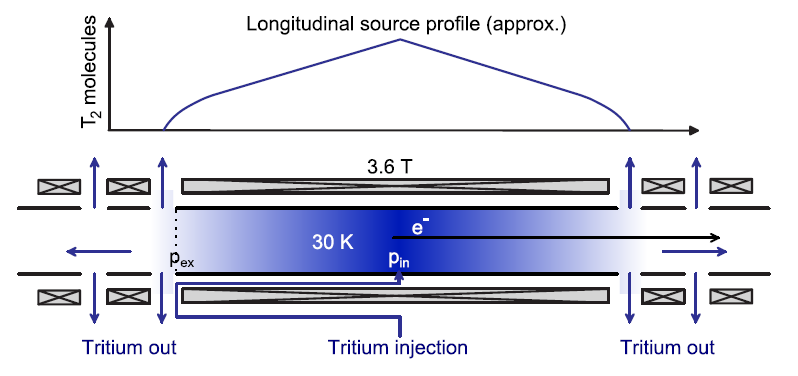}
        \caption{Setup of the WGTS. Tritium is injected into the centre of the cylinder and pumped out at both ends. The flux tube is surrounded by superconducting magnets to guide the $\upbeta$-electrons. The longitudinal density profile of the tritium molecules along the column is shown above.}
        \label{Figure:WGTS}
	\end{figure}

	The WGTS is the origin of $\upbeta$-electrons whose observed spectrum will ultimately lead to the measurement of the neutrino mass \cite{WGTS}. The general setup of the WGTS is shown in \fref{Figure:WGTS}. It is a column of tritium gas inside a cylinder with a diameter of \SI{90}{\milli\metre} and a length of \SI{10}{\metre}. The latter is situated in a homogeneous magnetic field of \SI{3.6}{\tesla} generated by superconducting solenoid magnets. The tritium gas is injected in the middle of the beam tube with an adjustable pressure $p_{\text{in}}$ = \SI{e-3}{\milli\bar}, and is pumped out at both ends with a constant outlet pressure of $p_{\text{out}} =$ \num{0.05} $p_{\text{in}}$. 
	
    \subsection{Column density}
	\label{Subsection:ColumnDensity}
	
	The column density is defined as tritium  molecule density integrated along the central axis of the source, i.e., the number of tritium molecules per source cross section area. The neutrino mass measurement depends on the accurate description of inelastic scattering of electrons by the gas molecules inside the source. There are several key parameters of the WGTS that need to be kept stable with high precision in order to achieve a high sensitivity in the neutrino mass measurement. These include
	\begin{itemize}
        \item \textbf{Beam tube temperature} \\
        The molecular tritium gas must be at cryogenic temperatures of \SI{<80}{\kelvin} to minimise corrections to the electrons energy due to thermal movement of the decaying mother atoms. The cooling concept is based on a two-phase liquid neon thermosiphon \cite{GROHMANN2009413,GROHMANN20135}.
        \item \textbf{Pressure} \\
        The amount of tritium inside the source scales with the inlet pressure. Stabilisation is achieved using a pressurised control vessel from which tritium flows via a capillary to the beam tube.
        \item \textbf{Tritium purity } \\
        A high isotopic purity of molecular tritium gas (\SI{>95}{\percent}) is required. The tritium purity $\mathbf{\epsilon_{T}}$ is given by the ratio of the number of tritium atoms to the total sum of atoms in the WGTS. In addition to T$_{2}$ other isotopolouges include DT, HT, D$_{2}$, HD, and H$_{2}$. The tritiated hydrogen isotopolouges differ in their mass, recoil energies, and the rotational and vibrational final state distributions of their daughter molecules following tritium decay. The gas composition is measured via \textit{LAser RAman spectroscopy} (LARA) \cite{LARA1,LARA2}.
	\end{itemize}
	These key parameters have an effect on the rate and/or energy of the electrons emitted from the source. There are several control and monitoring systems in the KATRIN experiment with the purpose of meeting the precision and stability requirements of the key source parameters.
	
	The column density, $\mathcal{N}$, can be obtained by combining an in-situ measurement of the tritium purity with an activity (decay rate) measurement. The count rate $S$ of $\upbeta$-electrons from the source as measured by dedicated particle detectors (activity monitors, see section \ref{Subsection:ActivityDetectors}) scales as
	\begin{equation}
        S = C \cdot \epsilon_{T} \cdot \mathcal{N}
        \label{Equation:ColumnDensity}
	\end{equation}
	where $C$ is a proportionality constant encompassing experimental properties such as detector efficiency and acceptance, and the half-life of tritium. Small fluctuations of the source parameters lead to changes in the observed shape of the differential $\upbeta$-electron spectrum. Fluctuations in the column density are expected to be in the \num{e-3} regime. Given the targeted sensitivity for the neutrino mass measurement, column density and tritium purity must not give rise to an uncertainty beyond $\delta m^{2}_{\nu} =$ \SI{7.5e-3}{\square\electronvolt} to the neutrino mass analysis. 

    \subsection{Electron transport}
	\label{Subsection:ElectronTransport}
	
	The $\upbeta$-electrons resulting from the decay of the tritium are adiabatically guided towards the spectrometer and detector section. The transport section is also used to eliminate the tritium flow towards the spectrometers which must be free of tritium in order to meet the necessary background requirements for neutrino mass measurements. The transport section consists of a \textit{Differential Pumping Section} (DPS) and a \textit{Cryogenic Pumping Section} (CPS).
	
	The DPS consists of five beam tube segments within superconducting solenoids with turbomolecular pumps between each \cite{MARSTELLER2021109979}. The CPS traps all remaining traces of tritium by cryo-sorption on argon frost at \SI{4}{\kelvin} condensed on the gold plated surfaces of the beam tube \cite{CPS2010,R_ttele_2017}. Both the DPS and CPS have \SI{20}{\degree} chicanes to block the line of sight for the diffusing tritium gas and to increase the probability that the tritium molecules get pumped away or hit the walls of the beam tube.
	
	At the end of the transport section the tritium flow is suppressed by \num{14} orders of magnitude compared to the center of the WGTS. The electron flow is unaffected and all electrons are guided adiabatically towards the spectrometer and detector section.


    \subsection{Activity monitors}
    \label{Subsection:ActivityDetectors}

	The activity of the tritium source is monitored by two systems. These activity monitors
	\begin{enumerate}
        \item provide information about fluctuations of the WGTS activity on a timescale of minutes and
        \item are used (together with the measured tritium purity) to monitor the column density with \SI{0.1}{\percent} precision, via \eref{Equation:ColumnDensity}.
	\end{enumerate}
     
    One of these activity monitors is located at the rear wall upstream of the WGTS. This detector measures the X-rays created when the $\upbeta$-electrons impact on the rear wall \cite{Babutzka:2012}. The second activity monitor is called the \textit{Forward Beam Monitor} (FBM). It is located downstream towards the main-spectrometer in the transport section, mounted between the last two superconducting solenoids of the CPS. Here the tritium flow has been suppressed by a factor of $10^{\num{14}}$, to approximately \SI{e-14}{\milli\bar\litre\per\second}, which minimises background effects and contamination from tritium. The magnetic field in this position is axially symmetric with a magnitude of \SI{0.84}{\tesla} so the spatial homogeneity of the source profile can be studied. The FBM is the final measurement component before the spectrometer and detector section.

	\section{Measurement principle}
	\label{Section:MeasurementPrinciple}
	
	The FBM measures $\upbeta$-electrons from the tritium source as they are guided to the spectrometer and detector section. Hence, the $\upbeta$-electrons are following the beamline when they are detected by the FBM. Such a detector must not shadow any part of the electron flux tube that will be used for the measurement of the neutrino mass. Therefore, the FBM configuration is such that the detector is located in the outer rim of the electron flux during neutrino mass measurements. The active radius of the flux tube used for measurement is approximately \SI{71}{\milli\metre} and the outer rim in which the detector is situated is up to \SI{7}{\milli\metre} wide. 
	
	The \pin diode detectors have an energy threshold of approximately \SI{5}{\kilo\electronvolt}, dependent on the background noise and the type of diode used. This lower energy value is determined during calibration of each diode. For an accurate rate measurement the lower energy threshold needs to be stable. 
	
	\begin{figure}[!ht]
        \centering
        \includegraphics[width=0.85\textwidth]{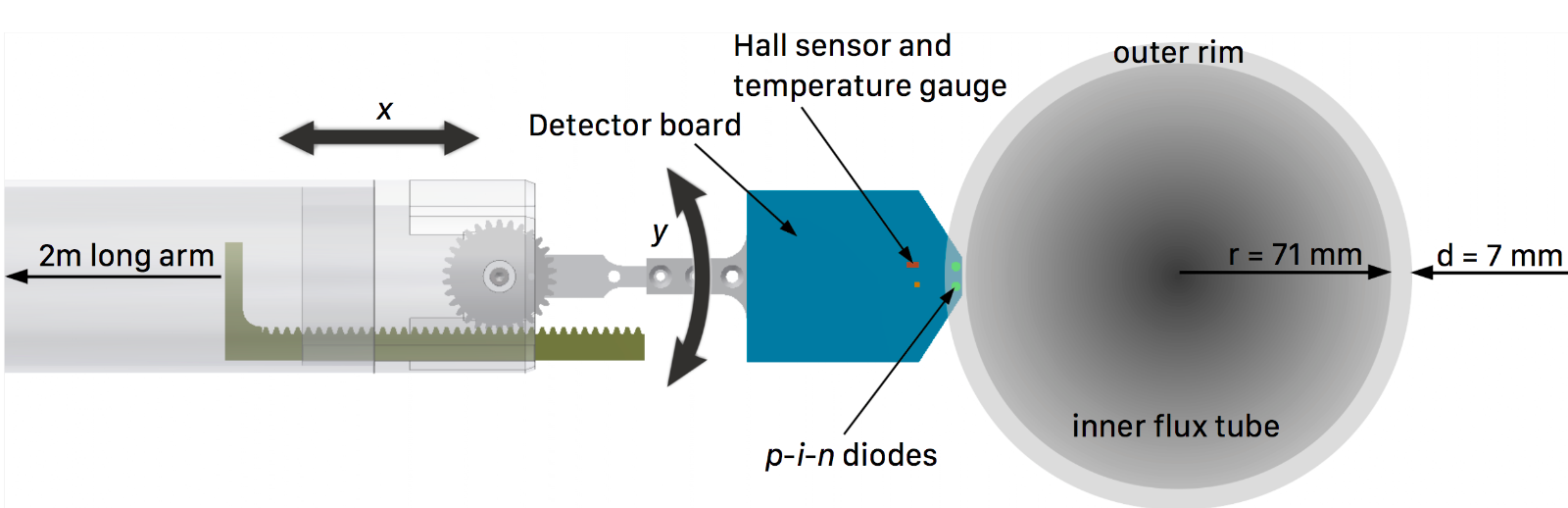}
        \caption{Cross section of the FBM setup with the electron flux tube in the KATRIN beamline. During nominal monitoring operation the FBM is situated in the outer rim of the flux tube, up to approximately \SI{7}{\milli\metre} in thickness.}
        \label{Figure:Fluxtube}
	\end{figure}
	
	It is assumed that the activity measurement in the outer rim of the flux tube is representative of the activity across the entire beamline cross section. Variations of the column density in the radial direction are expected to be on the \num{e-4} level~\cite{Hotzel:2012}. The assumption that the outer rim is representative of the entire flux tube is verified during repeated calibration runs when the FBM is moved across the beamline. These two operation modes of the FBM are standard ``monitoring mode'' and calibration ``scanning mode'' and are described in the following sections.
    
	\subsection{Monitoring mode}
	\label{Subsection:MonitoringMode}
	
	Monitoring mode is the standard mode of operation for the FBM. It is intended for permanent and continuous monitoring of the source activity and the main observable is the electron count rate. Together with the measurement of the tritium purity, the FBM monitoring mode provides continuously information on the column density of the source.
	
	\subsection{Scanning mode}
	\label{Subsection:ScanningMode}
	
	Flux tube scans are performed during calibration of the KATRIN experiment. The purpose of scanning is to
	\begin{enumerate}
        \item confirm that the activity in the beamline outer rim is representative of the entire flux tube,
        \item map any irregularities in the cross section of the flux tube, and
       
        \item investigate the area of the flux tube entering the spectrometer and detector section (i.e. measure possible shadow effects by STS instrumentation).
	\end{enumerate}
	During the KATRIN experiment calibration runs are performed between neutrino mass measurement runs once every \SI{\sim60}{days}. During commissioning and initial measurement campaigns the scanning mode was used more frequently.
	
	\section{Technical description}
	\label{Section:TechnicalDescription}
	
	In the following sections a technical description of the FBM is given. A more detailed description can be found in ref. \cite{Ellinger:2019}. Further information on the basic concept and the early development of the FBM can be found in ref. \cite{Schmitt:2008} and \cite{Babutzka:2010}.
	
	\subsection{Vacuum manipulator}
	\label{Subsection:VacuumManipulator}
\begin{figure}[!ht]
        \centering
        \includegraphics[trim = 0mm 290mm 0mm 0mm, clip, width=0.8\textwidth]{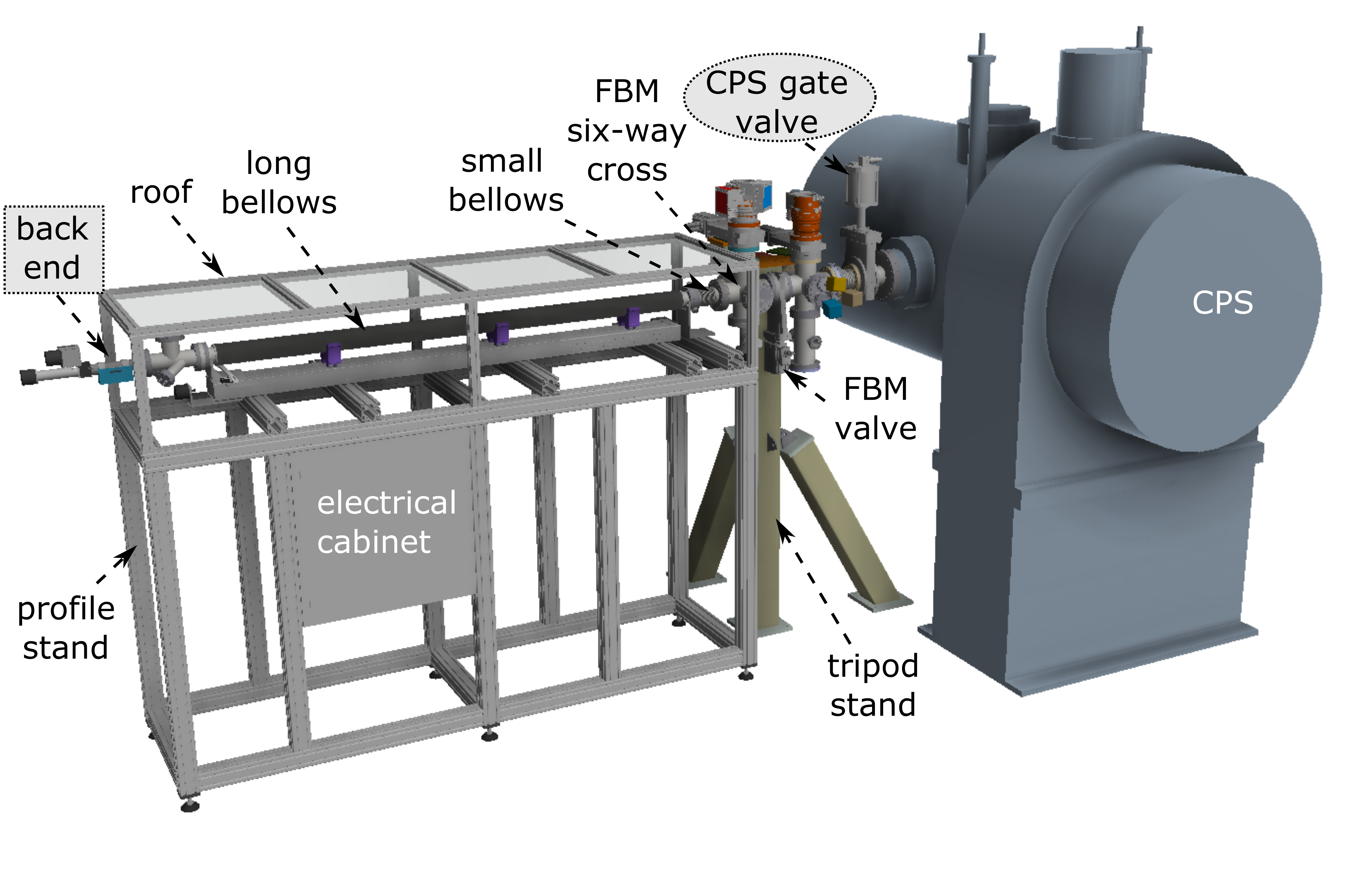}
        \caption{The FBM hardware setup. The FBM is attached to the CPS perpendicular to the beamline. The CPS and FBM gate valves separate the FBM from the CPS if the detector is in parking position within the FBM six-way cross. With the help of the \SI{2}{\metre} long bellow the detector arm can be driven into the flux tube within the CPS.}
        \label{Figure:Setup}
	\end{figure}

\begin{figure}[!ht]
        \centering
        \includegraphics[width=1.\textwidth]{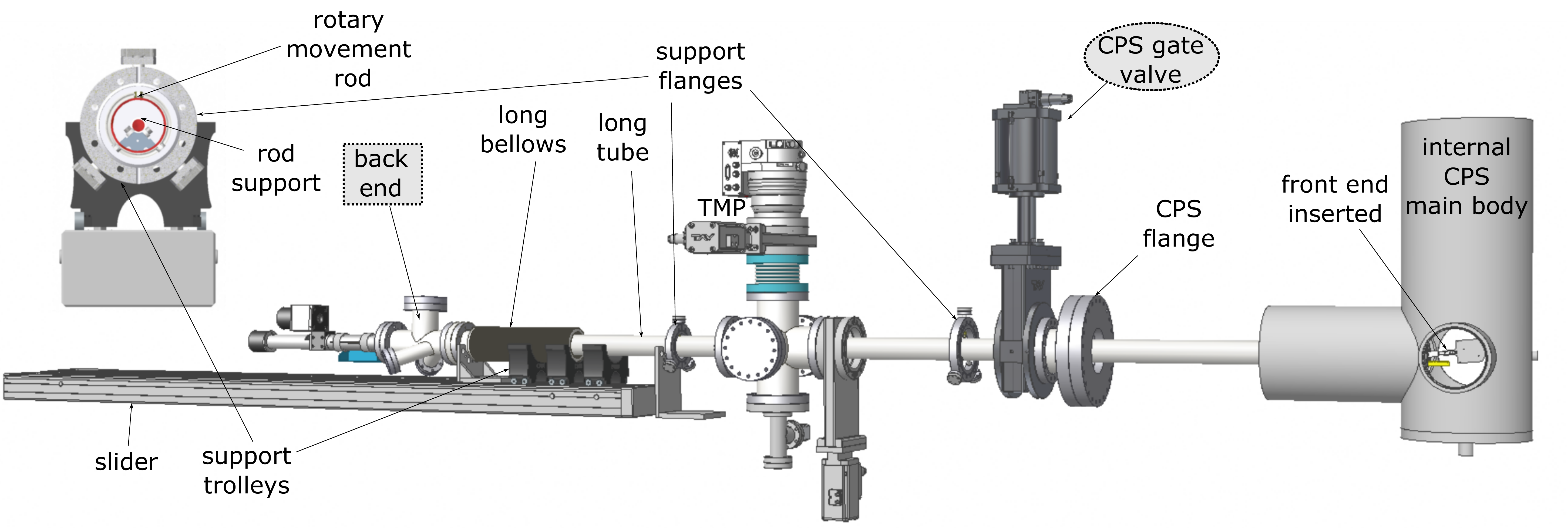}
      
        \caption{CAD drawing of the FBM as it is inserted into the CPS. In contrast to figure \ref{Figure:Setup} parts of the vacuum equipment, of the CPS and half of the FBM's long bellows are invisible for a better illustration of the mechanics.}
        \label{Figure:Setup_cut}
	\end{figure}

	The measurement of the electron flux is performed under ultra high vacuum (UHV) conditions in a potentially tritium contaminated environment.
	The main mechanical requirements for the vacuum manipulator are:
	\begin{enumerate}
        \item to situate the FBM detector board in the outer rim of the flux tube without shadowing the main detector and additionally to move it throughout the cross section of the flux tube,
        \item to be capable of removing all FBM components out of the CPS allowing full metal sealed vacuum gate valves to separate the FBM volume from the CPS volume, and
        \item to provide a safe enclosure for tritium, complying with all radiation safety regulations of the tritium laboratory.
	\end{enumerate}
	An overview of the complete FBM setup is shown in \fref{Figure:Setup} and \fref{Figure:Setup_cut}. 
	
    The vacuum components of the FBM setup are separated from the CPS by a gate valve. Behind this valve the FBM detector board is completely removed from the KATRIN beamline. Attached to this volume are a turbomolecular pump and pressure gauges. Behind the main FBM vacuum volume are bellows, support structures, stepper motors, rotary encoders, and electrical feedthroughs. These components provide the movement of the FBM detector board and the readout of the measured data.
	
\begin{figure}[!t]
        \centering
        \includegraphics[width=0.6\textwidth]{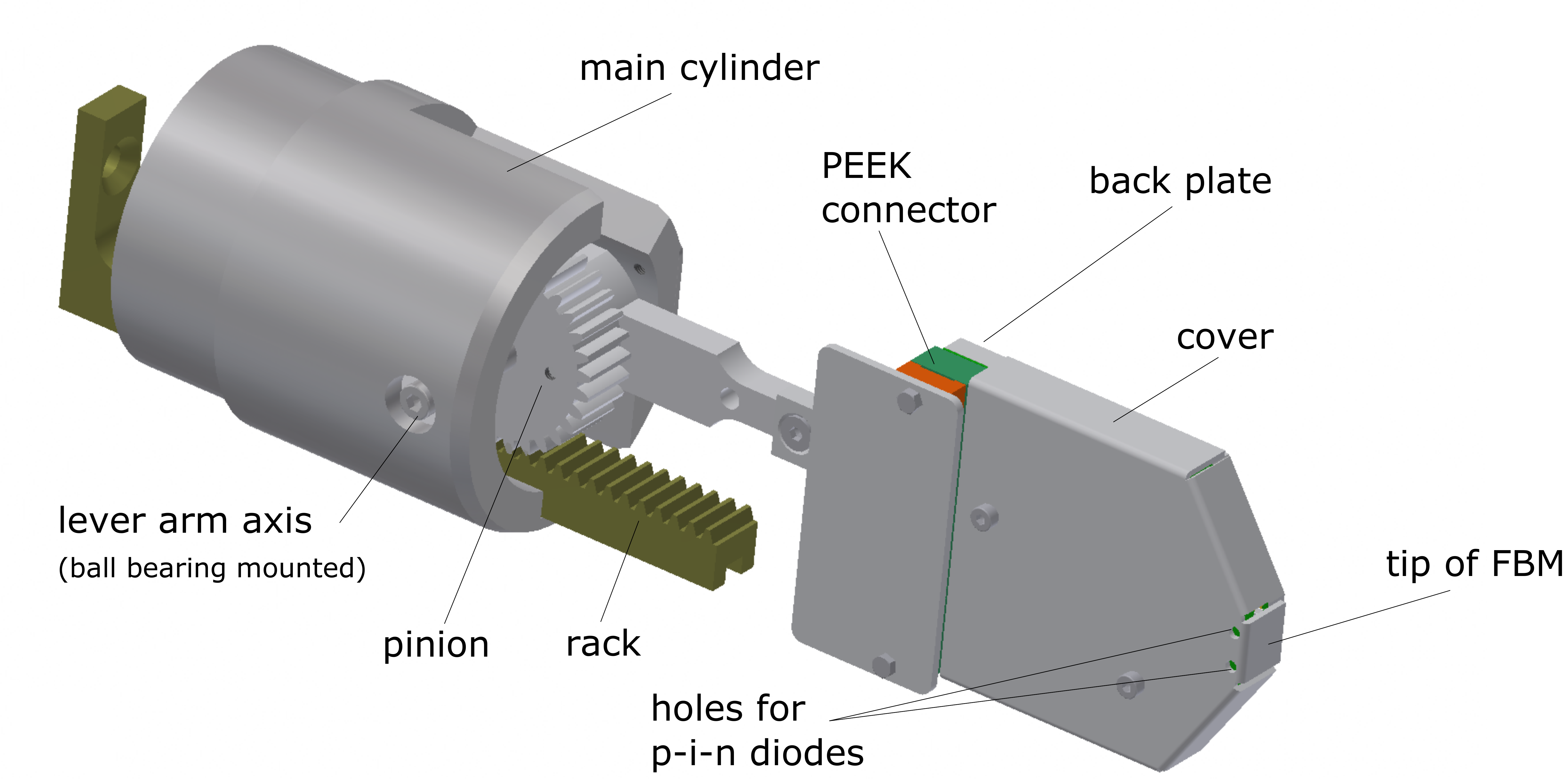}
        \caption{The FBM manipulator front end. The detector board is fixed on the end of a lever arm which is rotated by a rack and pinion drive.}
        \label{Figure:Frontend}
	\end{figure}

	The movement of the detector board  is realised by combining two linear drive mechanisms. A long stainless steel support tube with an outer diameter of \SI{54}{\milli\metre} can be moved over a distance of \SI{1.8}{\metre} along its symmetry axis. At its forward end the detector holder (hereafter known as the ``front end'', see figure \ref{Figure:Frontend}) is attached. The support tube provides space for electrical feeding and a driving rod which can be moved coaxially along the tube by approximately \SI{10}{\centi\metre}. The latter linear movement is converted by the front end into a rotary movement with a rack and pinion drive such that the combination of these two movements enable the positioning of the detector board in a two-dimensional plane. 
	
    Two edge-welded bellows are used to realise the linear movements in the vacuum. The large bellow has an unusually long extended length of \SI{2223}{\milli\metre} with a working-stroke of \SI{1800}{\milli\metre}. The back end provides electrical feedthroughs as well as the mechanics for the rotary movement. The system is moved with a \SI{2}{\metre} long spindle drive featuring low play and  two carriages for more stability. To prevent the long bellows, the driving rod, and the support tube from sagging and hanging down, several supports are added to the setup. These include 3D printed trolleys outside the vacuum which can move freely over the slider and are automatically pulled along from the motion of the bellows, and structures with full ceramic ball bearings for supporting the long tube and driving rod inside the vacuum chamber.

    The front end which contains the FBM detector board is the mechanical and electrical connection between the detector board and the manipulator. It converts the linear movement of the driving rod into a rotary movement with a low play rack-and-pinion drive to allow the movement in the vertical direction. It is shown in \fref{Figure:Frontend}.

    To reduce magnetic force acting on the system, as well as to reduce influences onto the electron guiding magnetic field, the front end, similar to all other vacuum parts of the FBM, is made of low permeability $\mu_{\text{p}}$  materials (such as stainless steel \num{1.4429} with $\mu_{\text{p}}$ \num{<1.005}).
    
    To prevent cold welding of moveable parts the materials of the pinion (stainless steel), rack (titanium), and the front end's cylindrically shaped main body (stainless steel), are alternated. A precise groove in the main cylinder allows leading the rack with low play. To reduce friction, an ultra low friction and UHV compatible dry lubrication is added, which mostly consists of a coating with tungsten disulfide.
    
    To facilitate an easy slipping onto the second support flange the cylinder has a chamber at its forward end. Two cut-outs extend the movement limits in $y$-direction and provide space for the electrical feeding. 

    The axis of the detector holder is made of steel \num{1.4429} like the pinion and is mounted via dry full-ceramic ball bearings. The lever arm is also made of steel \num{1.4429}, but the detector board holder (back plate) of aluminum, to reduce weight. To shield the detector board from radio frequency and, even more importantly, from the electron beam, a steel \num{1.4429} cover was designed featuring two small holes which allow electrons to reach the \pin diode chips. The full lever arm length from the axis to the tip of the FBM (including the cover, compare figure \ref{Figure:Frontend} ) is \SI{130}{\milli\metre} and the maximum height of the detector equals the height of the cover which is \SI{50}{\milli\metre}. The electrical connector is shielded from the electron beam by a thin steel plate.

    The turbomolecular pump is located vertically above the main FBM vacuum volume and is capable of pumping speeds up to \SI{260}{\litre\per\second} (nitrogen). Two pressure gauges are mounted below the FBM vacuum volume which cover the range from \SI{1.3e-10}{\milli\bar} to \SI{1.3e-2}{\milli\bar}. In order to reach the required vacuum level the setup is baked out periodically after being exposed to atmosphere.
	
	\subsection{Motion control}
	\label{Subsection:MotionControl}
	
	The two stepper motors mentioned in sub\sref{Subsection:VacuumManipulator} (\SI{12.1}{\newton\metre} and \SI{2.7}{\newton\metre} holding torque, \SI{1.8}{\degree} resolution) are not directly acting on the spindle axes but with one stage transmissions using toothed wheels. Since the FBM is not equipped with motor breaks the $x$-transmission is chosen such that the torque at the motor is sufficiently small to withstand the vacuum forces even if it is not powered anymore.

    Since it is possible that the stepper motors miss steps without being noticed, absolute rotary encoders are used to determine the position of the FBM because they retain the full information of the position even during a power cut. These optical encoders work with up to \num{16}-bit single turn and \num{14}-bit multi turn resolution, i.e. $2^{16}$ steps per revolution can be counted. This sums up to an overall resolution of $2^{30}$ steps. To minimise mechanical play both encoders are connected directly to their corresponding spindle axes. The main spindle has a slope of \SI{2.5}{\milli\metre}, hence a theoretical precision of \SI{e-5}{\micro\metre} can be reached. However, due to mechanical tolerances the actual precision is significantly lower as will be described in sub\sref{Subsection:Alignment}.

    To fulfill stringent safety requirements the motion control of the FBM is implemented on a \textit{Field-Programmable Gate Array} (FPGA) which continues to run during power cuts with the help of three backup accumulators. It directly monitors and controls the motor, encoders, and sensors and also includes a fast full safety retraction of the FBM which allows closure of the safety gate valves to separate the FBM volume from the CPS.
    
    The FPGA communicates with two KATRIN internal database systems: the \textit{ZEntrale datenerfassung Und Steuerung} (ZEUS) server and the \textit{Advanced Data Extraction Infrastructure} (ADEI) server~\cite{Kleesiek:2014}. All data obtained by the FPGA is automatically transferred and available on both servers. Safety-critical systems, such as vacuum pumps, valves, pressure gauges, and end switches, are integrated within the KATRIN PCS7 safety system.

	\subsection{Detector}
	\label{Subsection:Detectors}
	
	The main tasks of the FBM are to monitor the electron flux within the electron beam and to obtain the beta spectrum of tritium. Detector chips with a thin entrance window (dead layer) are used to allow the detection of electrons with energies below \SI{10}{\kilo\electronvolt}. In addition this also allows detection of low energy (\SI{<60}{\kilo\electronvolt}) photons which is important for calibrating the detector. The FBM features a UHV compatible two channel detector board, including detector chips of silicon type and additional sensors, as described below.
	
    \begin{figure}[!t]
        \centering
        \includegraphics[width=0.4\textwidth]{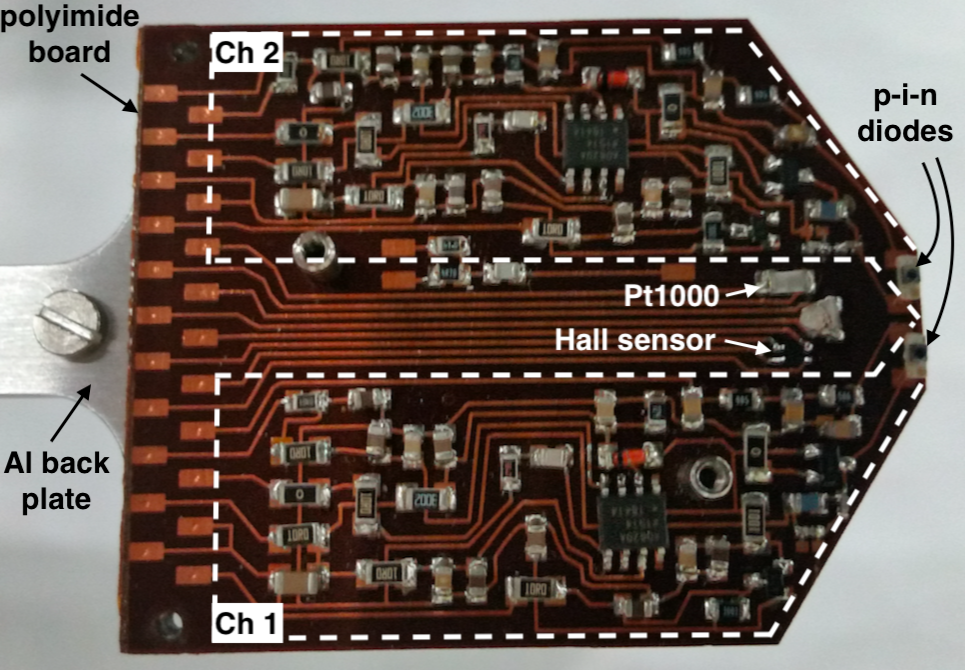}
        \hspace{0.8cm}
        \includegraphics[height=0.21\textwidth]{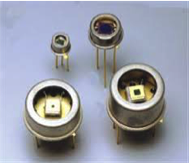}
        \includegraphics[height=0.21\textwidth]{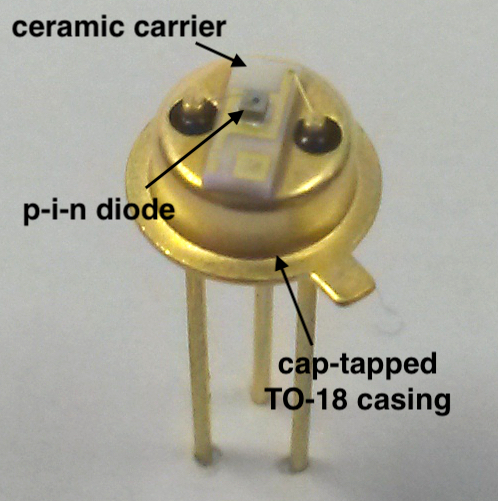}
        \caption{\textbf{Left:} The FBM detector board is made of polyimide and equipped with SMD parts. The two \pin diodes are glued to the tip of the board and their signals are amplified by two separate transimpedance amplifiers. Close to the \pin diodes a PT-1000 temperature sensor and a Hall sensor are located. \textbf{Center:} Collection of  \pin diodes in TO-18 casing as they are delivered by the producer \cite{Hamamatsu}.  \textbf{Right:} Picture of the Hamamatsu S9055 \pin diode with the lid removed. The silicon diode itself is mounted on a ceramic carrier which can be taken out of the casing.}
        \label{Figure:PINdiode}
	\end{figure}
	
    \subsubsection{Detector board and back plate}
    The detector board (PCB) is made of polyimide which meets the vacuum and material requirements because of its relatively low outgassing rate compared to most other polymers and excellent thermal resistance. To enhance the thermal through-plane conductivity of the board for dissipating the heat produced by the electrical components, the PCB is a flexible, thin (\SI{0.2}{\milli\metre}) multilayer board which consists of alternating polyimide and copper layers. Here the back plate acts as a passive heat sink which is permanently cooled in the cold environment of the CPS. The board contains two detector channels, each consisting of a detector chip and a preamplifier.

    A Hall sensor on the detector board determines the local magnetic field. 
    In this region of the CPS the magnetic field is approximately \SI{0.84}{\tesla} in the centre of the flux tube and is axially symmetric. The magnetic field is measured in only one axis and the electron flux should follow this magnetic field exactly with the exception of upstream blockages. The measurement of the magnetic field is therefore also useful for additional positioning and alignment measurements.
	
	Temperature stabilisation is important as the \pin diode leakage current rises exponentially with detector temperature. Therefore, the energy resolution and stability of the energy threshold are dependent on the detector temperature and effect the spectra obtained. To record the temperature a PT-1000 sensor is placed on the detector board near the \pin diodes and the Hall sensor.

    The PCB is mounted on a \SI{5}{\milli\metre}-thick aluminum back plate attached to the moving components. It is glued to the back plate with a UHV compatible two-component adhesive to ease the mounting of the electrical parts and to improve the thermal contact. The electronics are covered by a stainless steel metal shield to protect them from electrons and ions in the beamline as well as from radio frequency interference. To allow electrons to reach the \pin diodes the cover features two holes. The detector board has ``cut out'' corners in order to reduce the area of the flux tube that is covered. The electronics and detectors on the FBM detector board are connected via a custom-made PEEK connector with cabling running through the FBM manipulator to the vacuum feedthroughs.

\subsubsection{Preamplifiers and \pin diodes}
    The preamplifiers of the two \pin diode detector channels are DC coupled charge sensitive amplifiers which operate in a continuous reset mode. Each preamplifier consists of a low-noise JFET front end in common-source configuration and an operational amplifier (op-amp) connected in a non-inverting scheme. The feedback loop stretching across both stages consists of a  $R =$ \SI{1}{\giga\ohm} resistor in parallel with a $C =$ \SI{0.5}{\pico\farad} capacitor, forming a time constant of $\tau = C \cdot R =$ \SI{0.5}{\milli\second}. Thanks to the DC coupled circuitry, not only individual charge-generating events can be read out with a $F_{AC} = U/Q = 1/C =$ \SI{2}{\volt\per\pico\coulomb} translation factor, but also a current readout can be performed by looking at the DC voltage offset at the output of the preamplifier with $F_{DC} =$ \SI{1}{\volt\per\nano\ampere}.
    
    The fundamental components of the FBM are the \pin diode detector chips. There are two silicon \pin diodes mounted on the detector board which detect the $\upbeta$-electrons from the tritium source. These two \pin diodes can have different active sensitive areas. The silicon \pin diodes are manufactured by Hamamatsu Photonics and can be type S5971, S5972, S5973, or S9055-01 which have sensitive areas of different sizes (see table \ref{tab:pindiodes}). One advantage of these detectors is that their casing and properties are all identical, the only difference is their respective sensitive area. This means the electronic design of the detector board can remain the same and the board with the \pin diodes that most suits the measurement purposes can be mounted and inserted into the flux tube. Furthermore the dead layer does not exceed \SI{1}{\micro\metre}. 

    \begin{table}[!ht]
        \bigskip{}
        \centering%
        \begin{tabular}{cccc|cc|cc}
            \midrule
            \textbf{Diode} & \textbf{$\pmb{A_s}$ [\si{\milli\metre\squared}]}  & \textbf{$\pmb{C}$ [\si{\pico\farad}]} & \textbf{$\pmb{I_{\text{dark}}}$ [\si{\pico\ampere}]} & \multicolumn{2}{c}{$\pmb{d_{\text{dead}}}$ \textbf{[\si{\nano\metre}]}} &\textbf{$\pmb{T_{s}}$} \textbf{[\si{\second}]} & \textbf{$\pmb{T_{m}}$} \textbf{[\si{\second}]}  \\
            & \textbf{data sheet} & \textbf{(at \SI{10}{\volt})} & \textbf{(at \SI{10}{\volt})} & \textbf{data sheet} & \textbf{measured} & & \\
            \midrule
            S9055-01 & 0.008  & 0.5 & 2.0 & \multirow{5}{*}{<1000} 
                                          & \multirow{5}{*}{300--500}
                                              & 498  & 252  \\
            S9055    & 0.031  & 0.8 & 2.0 & & & 129  & 64.6 \\
            S5973    & 0.126  & 1.6 & 1.9 & & & 31.7 & 15.9 \\
            S5972    & 0.503  & 3.0 & 10  & & & 8.0  & 4.0  \\
            S5971    & 1.131  & 3.0 & 70  & & & 3.5  & 1.8  \\
            \midrule
        \end{tabular}
        \caption[Parameters of the FBM \pin diodes.]{Parameters of the FBM \pin diodes. Capacitance $C$ and dark current $I_{\text{dark}}$ are taken from the data sheets. The thickness of the dead layer $d_{\text{dead}}$ was determined with an electron gun and Monte Carlo simulations. The two right columns show the time to build a sufficiently detailed spectrum in monitoring mode ($T_{m}$) and scanning mode ($T_{s}$).}
        \label{tab:pindiodes}
    \end{table}
    
	The casing of these diodes is metal and includes a large glass window. Since the windows of these TO-18 casings would prevent the detection of any electrons the diodes are removed from the housing and directly mounted (using two-component adhesive) onto the FBM detector board. The Hamamatsu S5971 \pin diode detector chip is shown in \fref{Figure:PINdiode}.

	The choice of the \pin diode size is based on the expected rate from the tritium source within each measurement phase (larger diodes are used for commissioning measurements where the amount of tritium is lower).

    The statistical error of the measurement is dominated by the number of electrons that are counted by the detector and is given by
    \begin{equation}
        \frac{\Delta N}{N} = \frac{1}{\sqrt{N}} = \frac{1}{\sqrt{A\phi\epsilon t}} 
    \end{equation}
    where $A$ is the sensitive area of the \pin diode, $\phi$ is the electron flux density, $\epsilon$ is the detector efficiency, and $t$ is the measurement time. The detector efficiency includes losses due to back reflected electrons and pile-up effects. To reach the required precision of $\Delta N/N =$ \SI{0.1}{\percent} the measurement time is
    \begin{equation}
        t = \frac{1}{0.001^{2}A\phi\epsilon}
    \end{equation}
    Assuming an energy threshold of \SI{7}{\kilo\electronvolt} approximately $\frac{1}{3}$ of the tritium spectrum is measured. Using this reduction factor, an electron flux density of \SI{e6}{\per\second\per\milli\metre\squared} and a detector efficiency of $\epsilon =$ \SI{65}{\percent} the measurement time needed to reach the required \SI{0.1}{\percent} precision for each of these \pin diodes is calculated and listed in \tref{tab:pindiodes}.

    The one unknown property of these \pin diodes is their individual dead layer. During manufacturing the thickness of the dead layer is not measured and therefore not available a priori, but limited to \SI{1000}{\nano\metre}.
    The thickness of the dead layer is indicated by the minimum energy that can be detected. The measurement of the dead layer is done by analysing  the shape of the peak from monoenergetic electrons originating from an electron gun (see section \ref{Section:Detectorresponse}). Figure \ref{Figure:eGun} illustrates such an analysis. Measurements of the dead layer are performed for each \pin diode before they are mounted on the FBM detector board. It is assumed that the dead layer remains constant over time, even after bakeout cycles of the vacuum setup. This is because the dead layer is silicon oxide which is not affected by heat and requires approximately \num{e13} electrons (on the order of several years in the FBM location) to suffer from radiation damage.

	\subsection{Data acquisition}
	\label{Subsection:DataAcquisition}
	
	For the two \pin diode detector channels an Amptek \cite{Amptek} PX5 and an Amptek DP5 are used for the data readout. These are digital pulse processors with build-in amplifiers used to amplify the signal by up to a factor of \num{100}. These Amptek devices are connected to an Apple (Mac mini) computer running the \textit{Object-orientated Real-time Control and Acquisition} (ORCA) software \cite{Howe2004}. An ORCA readout module was specifically designed for the FBM Amptek devices. The raw ORCA data is converted into ROOT \cite{Root} files for analysis. The preamplifier outputs of the two \pin diode detector channels can also be connected to separate low-pass filters to measure the DC offset occurring from the event rate on the respective \pin diode chip.

	The pulse processing parameters of each detector channel can be optimised to obtain either the count rate or a spectrum of the $\upbeta$-electrons from the source. The peaking time is set to
	\begin{itemize}
        \item Fast channel: \SI{1.0}{\micro\second} to measure the count rate (larger \pin diode with higher count rate)
        \item Slow channel: \SI{3.2}{\micro\second} to measure the spectrum (smaller \pin diode with lower count rate)
	\end{itemize}
    During scanning the required measurement time at each point is reduced due to the increased electron flux towards the centre of the beam tube. The analysis of the FBM data is based on the established analysis systems of the KATRIN experiment. Therefore, all data, slow control, and run files are available on the ADEI server and KATRIN databases.

    \section{Measurements}
	\label{Section:Measurements}
	
	This section presents selected results \cite{Ellinger:2019} of the measurements performed with the FBM during its commissioning phases as well as during the first KATRIN measurement campaigns. These results serve as an evaluation tool for the positioning accuracy of the vacuum manipulator and the performance of the detector. In some cases the data is compared to the results of numerical simulations of the detector response.

	\subsection{Alignment and positioning precision}
	\label{Subsection:Alignment}
	
	\begin{figure}[!t]
        \centering
        \includegraphics[trim = 0mm 0mm 150mm 0mm, clip, width=0.42\textwidth]{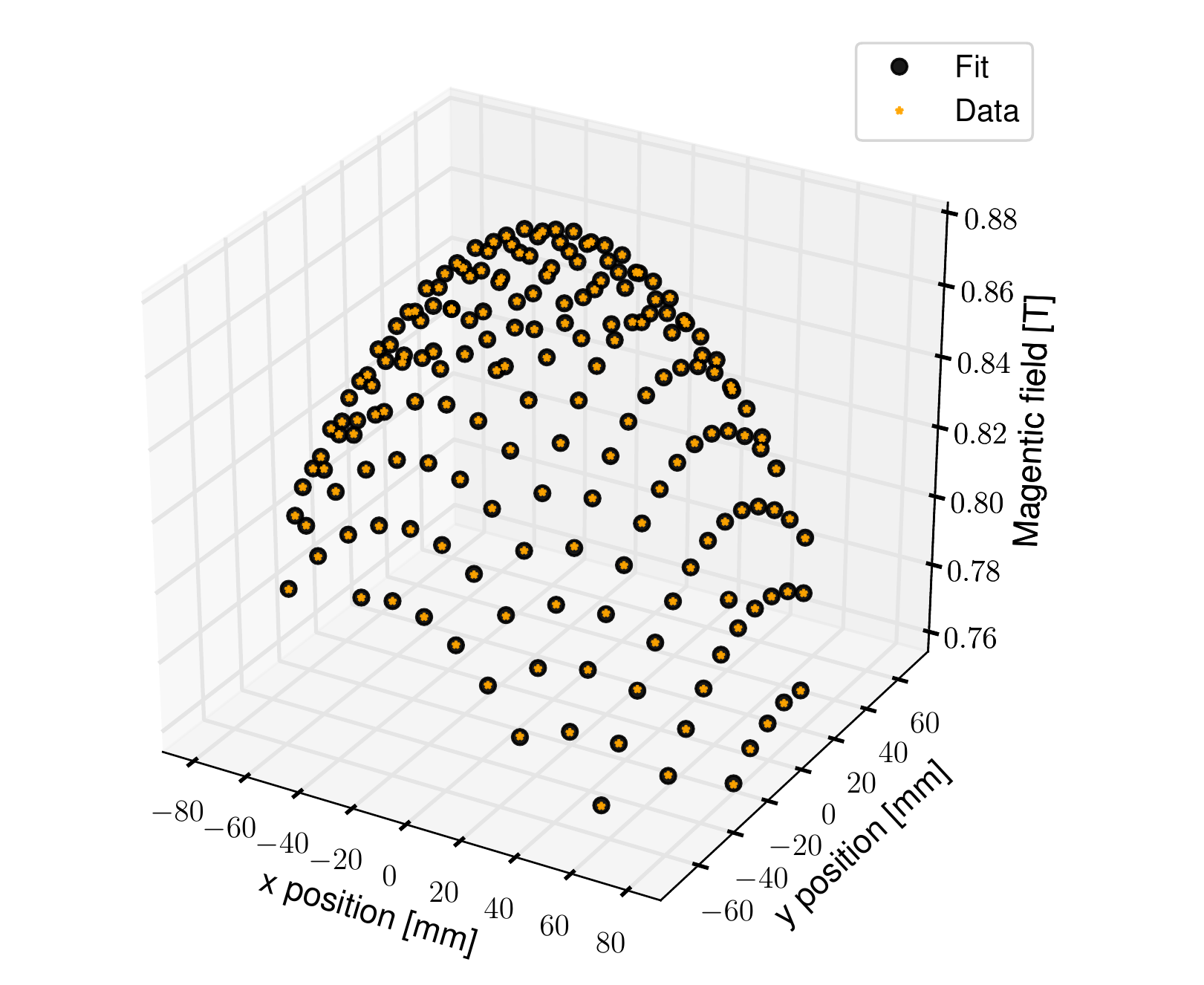}
        \includegraphics[width=0.42\textwidth]{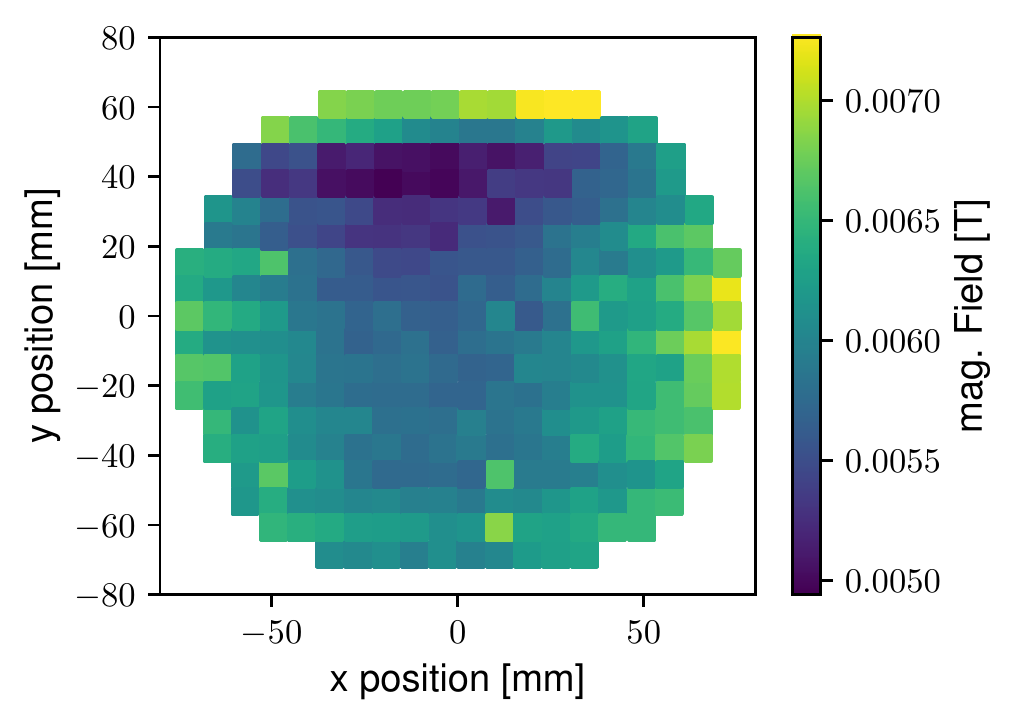}
        \caption{\textbf{Left:} Data and fit result of the calibrated and temperature corrected $z$-component of the magnetic field in the CPS. \textbf{Right:} The residuals of the simulated magnetic field which shows a good agreement with the data.}
        \label{Figure:Alignment}
	\end{figure}
	
	Positioning reproducibility is the ability of the FBM to find a position relative to a former position. This is different to the absolute positioning accuracy which includes external reference points with respect to the KATRIN coordinate system. The reproducibility is validated by using a laser setup as well as a portable \textit{Coordinate Measuring Machine} (CMM). It was determined to be better than \SI{0.1}{\milli\metre}. However, the overall alignment uncertainties (also CMM) dominate the absolute positioning accuracy as shown in table \ref{tab:overall_error}.
	
    \begin{table}[!ht]
        \centering
        \begin{tabular}{ccccc}
            \hline 
            & \textbf{$\pmb{\sigma_{\text{alignment}}}$ [\si{\milli\metre}]} &\textbf{$\pmb{\tilde{\sigma}_{\text{max}}}$ [\si{\milli\metre}]} &\textbf{$\pmb{\sigma_{\text{full}}}$ [\si{\milli\metre}]}& \textbf{Offset [\si{\milli\metre}]} \\
            \hline 
            $x$ & \num{0.28} & \num{0.042} & \num{0.28} & \num{-1.2\pm0.2} \\
            $y$ & \num{0.1}  & \num{0.07}  & \num{0.13} & \num{4.2\pm0.2}  \\
            \hline 
        \end{tabular}
        \caption{The overall positioning accuracy $\sigma_{\text{full}}$ results from the combination of the uncertainties of the alignment $\sigma_{\text{alignment}}$ and the positioning reproducibility $\tilde{\sigma}_{\text{max}}$. There is a misalignment between the FBM and the flux tube expressed by a constant offset which was determined from flux tube scans.}
        \label{tab:overall_error}
    \end{table}

    To calibrate the movement system, as well as to find the center of the flux tube, the magnetic field in the CPS can be used (see left panel in figure \ref{Figure:Alignment}). The shape of the magnetic flux can be described by a two-dimensional Gaussian. The required calibration values, namely the encoder value for the horizontal lever arm and the offset of the magnetic flux center to the FBM system (listed in the last row in table \ref{tab:overall_error}), are given by the free parameters in a fit of data taken during a flux tube scan.

\begin{figure}[!ht]
        \centering
        \includegraphics[width=0.40\textwidth]{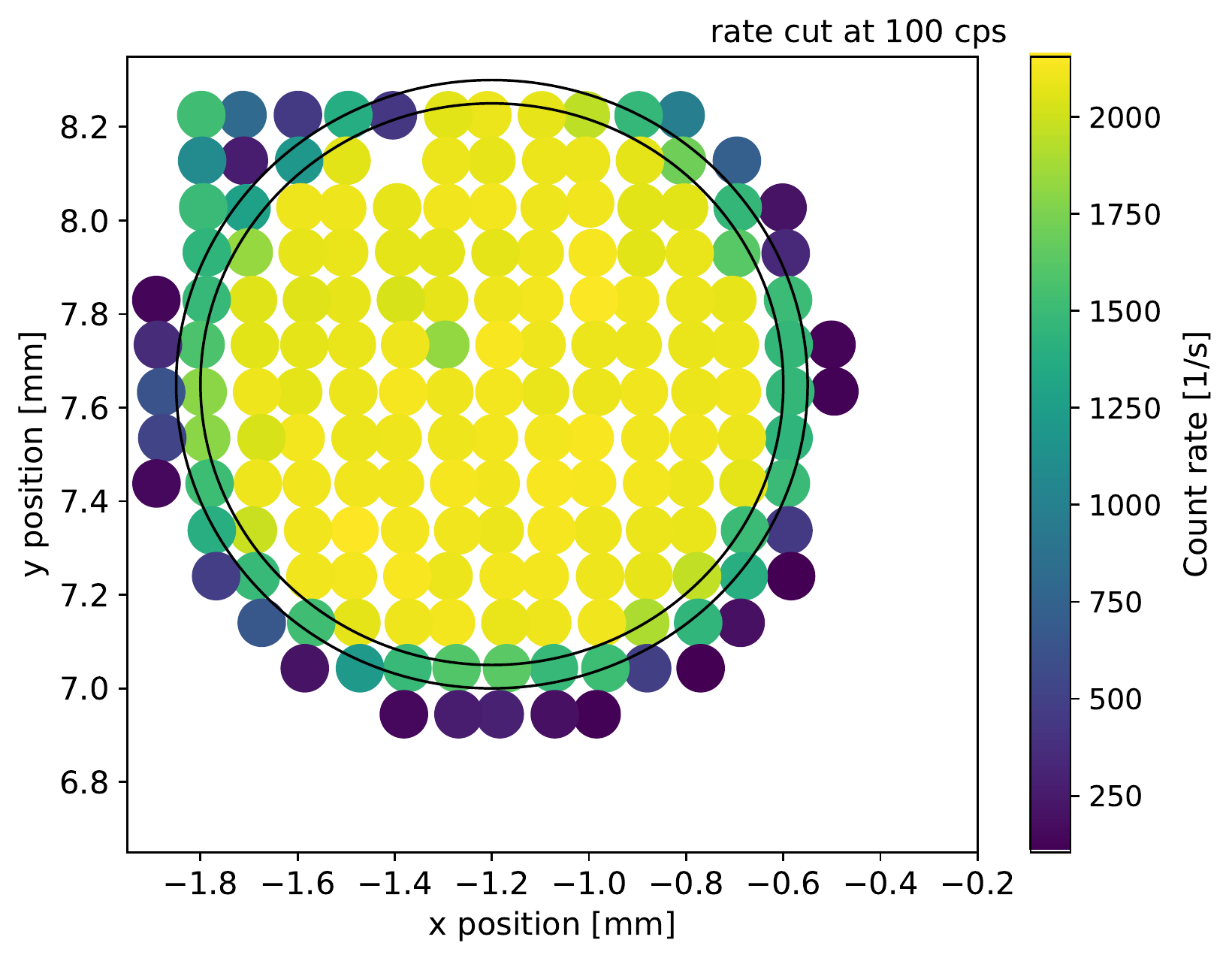}
        \caption{Scan of stationary electron gun beam. The FBM detector board is moved through the beam in a grid with \SI{0.1}{\milli\metre} step length. Each colored dot represents the measured rate at this detector position. The background is removed by a noise cut at \SI{100}{cps}. Note that the size of the data points is arbitrary and does not represent the size of the beam spot. The inner circle represents the active surface (here \SI{1.13}{\milli\metre\squared}) as stated by the manufacturer and the outer circle the visual surface of the \pin diode. The detector chip profiles are positioned such that they comprise the highest rate.}
        \label{Figure:eGun2}
	\end{figure}

    To demonstrate the excellent positioning accuracy of the manipulator a thin (\SI{0.14}{\milli\metre} diameter) electron beam was scanned with the FBM by moving the detector (type S5971 with \SI{1.2}{\milli\metre} diameter) through the fixed beam in a grid pattern with \SI{0.1}{\milli\metre} spacing \cite{Ellinger:2019}. Since the beam is far smaller than the \pin diode, it is rather the diode being scanned by the beam than vice versa. The plot in figure \ref{Figure:eGun2} shows the measured intensities as a function of detector position. The background is removed by a noise cut at \SI{100}{cps} and as a result only positions where electrons were detected are shown. The large circular contours represent the entrance window of the diode (small, \SI{1.2}{\milli\metre} diameter) as it is stated in the data sheets and the visual surface (large, \SI{1.3}{\milli\metre}) of the diode as it was measured. The position of the contours is adjusted such that the number of events within the contours is maximised. The center represents the actual position of the beam at $x_{\text{FBM}} =$ \SI{-1.2}{\milli\metre} and $y_{\text{FBM}} =$ \SI{7.6}{\milli\metre}. One spot close to the center has low statistics, which is associated with a small dirt particle which was found to be located at the equivalent position of the \pin diode chips's active surface.

	\subsection{Detector response and dead layer}
	\label{Section:Detectorresponse}
	
\begin{figure}[!ht]
        \centering
        \includegraphics[width=0.40\textwidth]{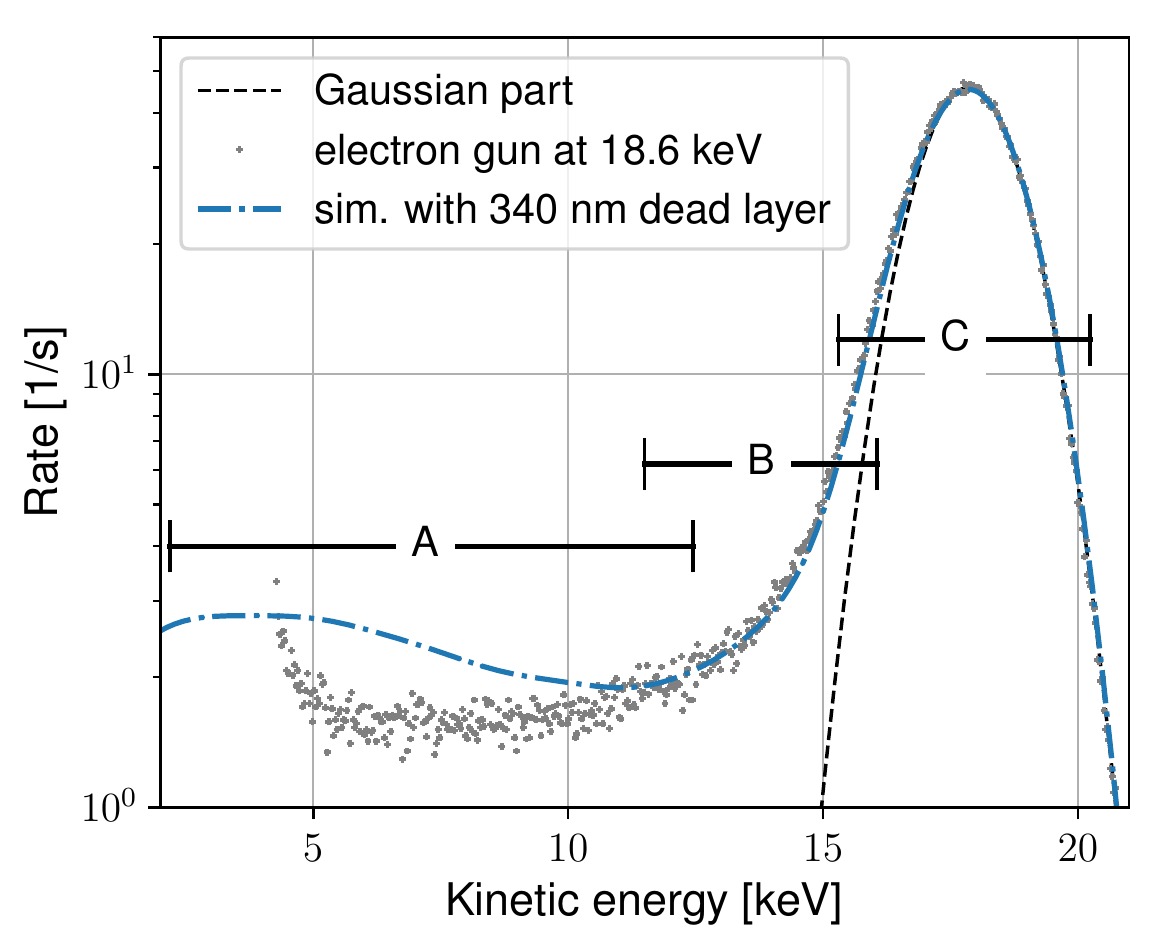}
        \includegraphics[width=0.445\textwidth]{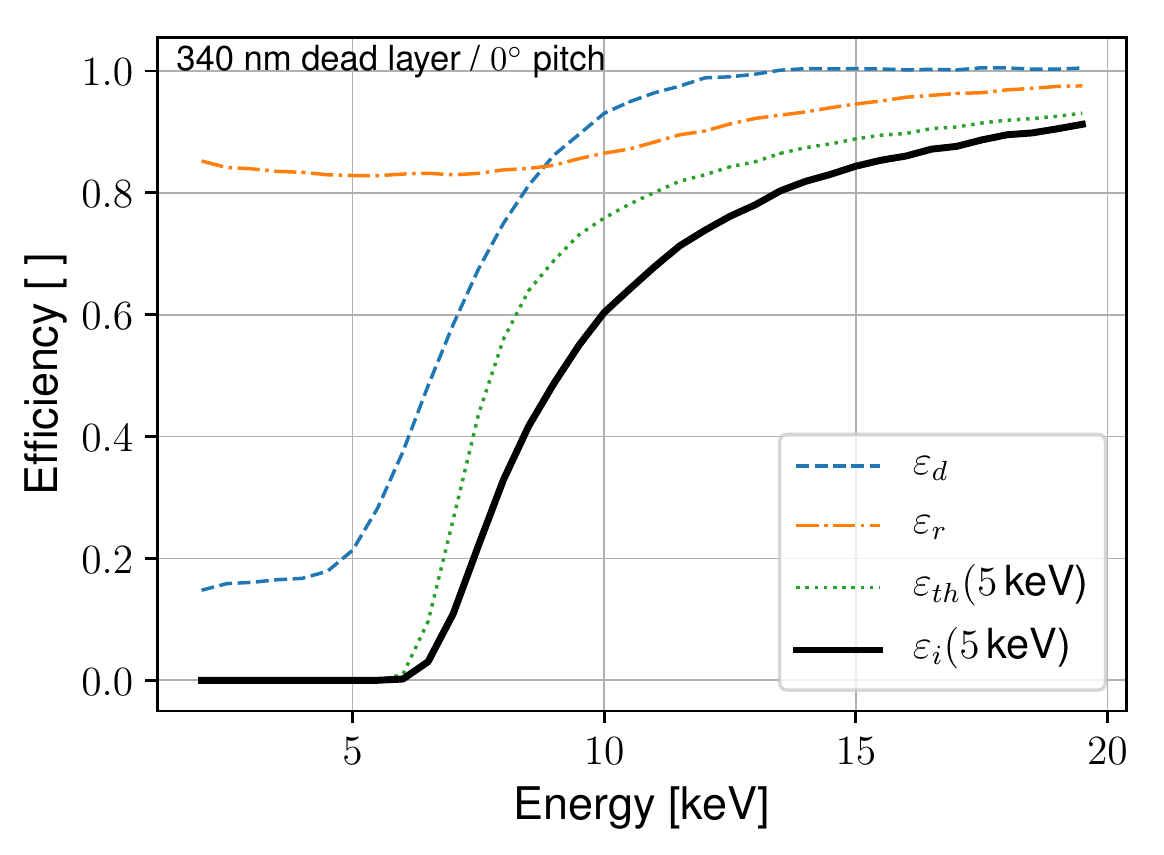}
        \caption{\textbf{Left:} Measured and simulated electron gun peaks obtained during the first tritium campaign. The simulation includes a detector energy resolution with FWHM $=$ \SI{2.35}{\kilo\electronvolt}. The best match was obtained with a dead layer thickness of \SI{340}{\nano\metre}. \textbf{Right:} The simulated efficiencies for electrons not to get reflected from the detector ($\epsilon_{r}$), not to get stopped in the dead layer ($\epsilon_{d}$), and for exceeding the energy threshold of \SI{5}{\kilo\electronvolt} ($\epsilon_{\text{th}}$ (\SI{5}{\kilo\electronvolt})). The intrinsic efficiency of the detector is then given by $\varepsilon_{i}$(\SI{5}{\kilo\electronvolt}) $= \varepsilon_{r}\cdot\varepsilon_{d}\cdot\varepsilon_{\text{th}}$(\SI{5}{\kilo\electronvolt}).}
        \label{Figure:eGun}
	\end{figure}

	For calibration KATRIN is equipped with an electron gun which is situated in the rear section and can provide a mono-energetic electron beam with energies up to \SI{20}{\kilo\electronvolt}. In the left panel of figure \ref{Figure:eGun} the measured detector response to \SI{18.6}{\kilo\electronvolt} electrons is shown. Due to partial charge collection the peak is shifted to lower energies, widens and develops a long low energy tail descending into an almost flat plateau. To understand the related effects and to reach the required precision for the FBM, numerical simulations \cite{Ellinger:2019} were performed using  Geant4 \cite{Agostinelli:2003}. The model of the detector comprises a detection volume and a fully insensitive dead layer (also known as slab-model). Within the detector the electrons undergo elastic and inelastic scattering losing their energy subsequently and are finally either completely stopped in the detector or reflected from it. For the interaction model the Penelope physics package was used. 
	The low energy region A (see figure \ref{Figure:eGun}) is predominantly produced by reflected electrons and region B mainly by the dead layer losses which also cause the shift of the main peak in region C by approximately $\SI{0.6}{\kilo\electronvolt}$ (mean energy loss in the dead layer).
	
	The measured electron spectra were compared to a library of simulated electron spectra for different dead layers, with a $\chi^2$ test used to find the most accurate match. The best result was obtained with a dead layer thickness of \SI{340}{\nano\metre}. The simulations always overestimated the data in the low energy tail at approximately  \SI{5}{\kilo\electronvolt} (region A). This is caused by an incomplete model which does not include the magnetic field configuration in the CPS for the purpose of simplicity. Due to magnetic mirroring in the CPS electrons which were initially reflected from the detector are guided back within the peaking time of the DAQ contributing to regions B and C instead of A.
	
	It was possible to determine the dead layers of the FBM \pin diodes which range from \SIrange{300}{500}{\nano\metre} causing an energy dependent shift of the measured peak of \SIrange{0.5}{2}{\kilo\electronvolt} for electron energies up to \SI{20}{\kilo\electronvolt}. With these simulations the detection efficiency for electrons as a function of kinetic energy could also be determined as shown in the right plot in figure \ref{Figure:eGun}.
	
	\subsection{First tritium measurement campaign}
	\label{Subsection:FirstTritium}

	\begin{figure}[!ht]
        \centering
        \includegraphics[width=0.70\textwidth]{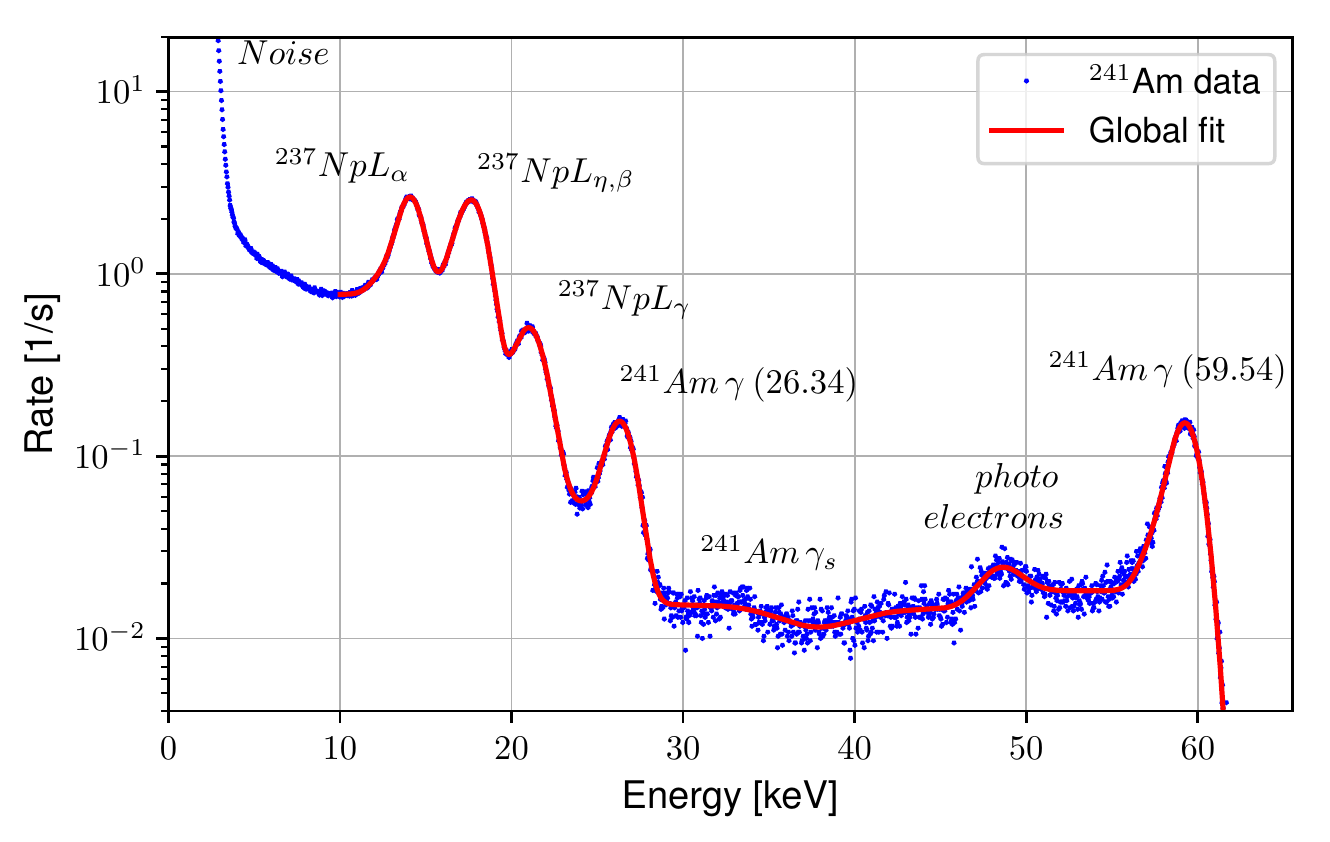}
        \caption{Data and fit of the $^{241}\text{Am}$ spectrum. The spectrum is recorded with the FBM with an energy resolution of $\sigma_{\text{FWHM}}\approx$ \SI{2}{\kilo\electronvolt}. The positions of all $^{241}\text{Am}$ lines are identified and labeled. The full fit function consists of a combination of Gaussians and error functions for each (strong) $^{241}\text{Am}$ line in the peak. The global fit comprises \num{33} free fit parameters. Only the Gaussian parts representing the strongest lines are used for the calibration of the FBM.}
        \label{Figure:AM241Calibration}
	\end{figure}
	
    Before the actual tritium measurement an alternative front end, equipped with a Faraday cup, was installed to the FBM in order to check ion blocking, measure the radial ion distribution in the beamline, and check the simulated source gas models by measuring secondary electrons \cite{PhDKlein2018}. The measurements with the \pin diode detector started with the ``first tritium measurement campaign`` \cite{Anker:2020} which took place from the \nth{5} to the \nth{20} of May 2018 with a gas mixture of \SI{0.5}{\percent}  tritium in deuterium. In the following sections the results of this first data-taking period with tritium are presented.
    
    \subsubsection{Configuration}
    \label{Subsection:Configuration}
    
    With a fraction of only \SI{0.5}{\percent} of tritium in the source gas an electron flux of approximately \SI{5000}{\per\second\per\milli\metre\squared} was expected at the FBM measuring plane. Therefore, the largest \pin diodes have been chosen (\SI{1.1}{\milli\metre\squared}) to optimise counting statistics. The peaking time of the DAQ for both channels was \SI{6.4}{\micro\second}, resulting in a pile-up rate of about \SI{3}{\percent} which can be neglected for stability analyses (see section \ref{Section:RateStability}). 
    
    Acceptance tests were performed prior to the campaign to extract calibration parameters, energy resolutions, and noise thresholds of the detectors. These measurements were performed with an $^{241}$Am source in the vented system with the FBM in parking position. The source was placed at a close distance between the two \pin diodes. The desired diode could then be irradiated using the movement mechanics and be adjusted to find the maximum count rate. In figure \ref{Figure:AM241Calibration} one of the $^{241}$Am spectra extracted from these measurements is shown. The calibration parameters are obtained by a global fit to the whole spectrum.
    
    \subsubsection{Spectrum}
    \label{Section:Spectrum}
    
\begin{figure}[ht]
        \centering
        \includegraphics[width=0.60\textwidth]{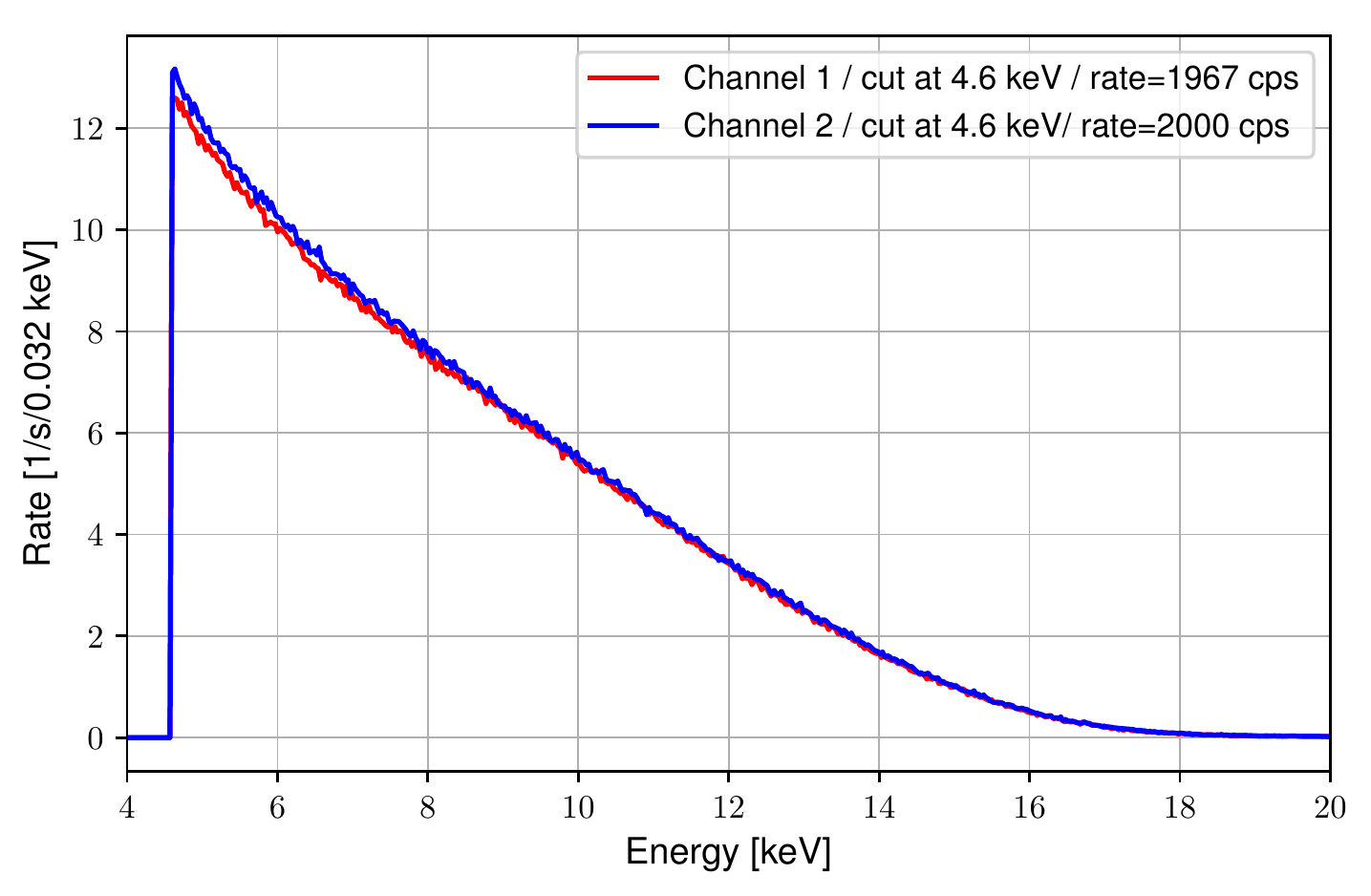}
        \caption{Tritium $\upbeta$-spectrum measured with both detector channels of the FBM during the first tritium campaign. The rate deviation of approximately \SI{2}{\percent} between the two channels is probably caused by the uncertainties in the calibrations or differences in the active surface or dead layer thickness of the \pin diodes.}
        \label{Figure:TritiumSpec}
	\end{figure}

    The spectrum shown in figure \ref{Figure:TritiumSpec} is the first tritium spectrum recorded with the FBM. Between \SIrange{6}{20}{\kilo\electronvolt} the spectrum agrees with the expectation, however below \SI{6}{\kilo\electronvolt} the slope is unexpectedly increasing. This is probably due to background counts from noise and edge effects from the diodes. This may also explain why the spectra of the two channels do not match perfectly for lower energies. Other likely sources for this mismatch, which is also the reason for about \SI{2}{\percent} lower rate in channel \num{1} than in channel \num{2} during the whole campaign, are 
    \begin{itemize}
        \item uncertainties in the energy calibrations which cause the deviations among the channels for lower energies, 
        \item small differences in the active area, or
        \item small differences in the dead layer thickness of the two \pin diodes.
    \end{itemize}
    
    \subsubsection{Flux tube scans}
    \label{Subsection:FluxTubeScans}
    
    Several scans of the $\upbeta$-electron flux cross section were performed recording the tritium count rate, the magnetic field, and the temperature. During a scan, the temperature usually drops by about \SI{1}{\celsius}. This occurs when the detector is moved further into the cold CPS where the detector directly faces the \SI{4}{\kelvin} cold beam tube of the CPS in which the argon frost layer is prepared. To check the calibration of the magnetic field sensor as well as to verify our understanding of the magnetic field configuration in the FBM measuring plane the magnetic field data is compared with simulations. The residual analysis displayed in figure \ref{Figure:Alignment} (right)  shows a good agreement between simulation and data.

    \begin{figure}[!ht]
        \centering
        \includegraphics[width=0.70\textwidth]{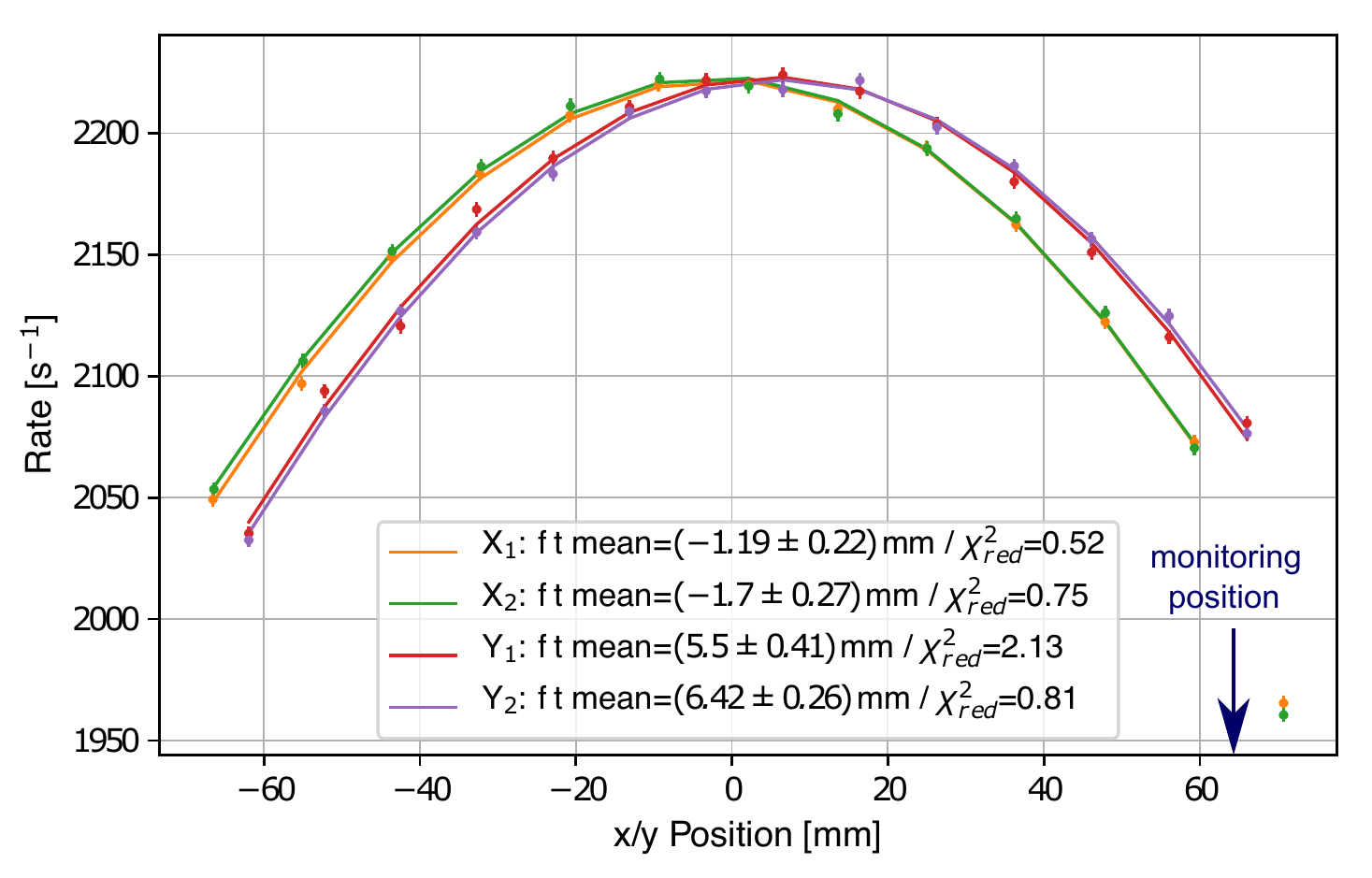}
        \includegraphics[width=0.70\textwidth]{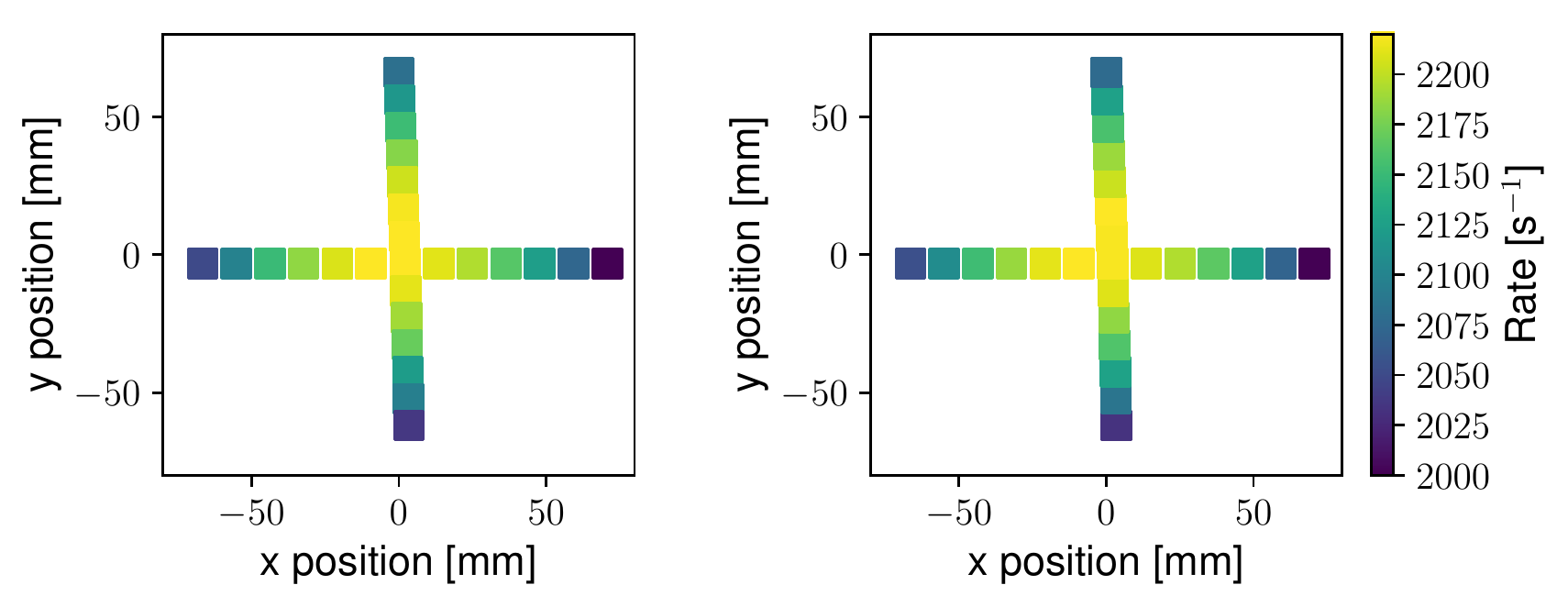}
        \caption[Radial dependence of the count rate derived from a cross scan during the first tritium campaign with channel \num{1} of the FBM.]{{\small{}Radial dependence of the count rate derived from a cross scan during the first tritium campaign with channel \num{1} of the FBM.} \textbf{\small{}Top:}{\small{} 1D Gaussians are fit to the data for each horizontal ($\text{X}_{1,2}$) and vertical ($\text{Y}_{1,2}$) scan. The Gaussian means are compatible with the results from magnetic field measurements. One can clearly see that for identical positions slightly different rates are measured, for example the rate increased during the $x$-scans such that the mean of the $\text{X}_{2}$ fit is lower than for $\text{X}_{1}$. The Gaussian widths are approximately $\sigma =$ \SI{165}{\milli\metre}.} \textbf{\small{}Bottom:}{\small{} 2D scatter plot of the same data. The scans for $y$ are not perfectly on a vertical line due to the chosen scan pattern which explains the larger uncertainties in the fits.}}
        \label{Figure:TritiumRateScan}
    \end{figure}

   In figure \ref{Figure:TritiumRateScan} (top and bottom) the results of a cross scans over the cross section of the flux tube for both detector channels are shown. The electron flux shows the expected Gaussian shape where the rate drops from the center to the outer rim by approximately \SI{10}{\percent} as predicted by simulations \cite{Haussmann:2019}. It can be seen that the event rate for identical positions changes during the scans which affects the extracted mean of the fits. Nevertheless, the means are compatible to the results from the alignment measurements in section \ref{Subsection:Alignment} which use the magnetic field data. This is expected as the electron flux scales with the magnetic flux.

    \subsubsection{Rate stability}
    \label{Section:RateStability}
    \begin{figure}[!ht]
        \centering
        \includegraphics[width=0.70\textwidth]{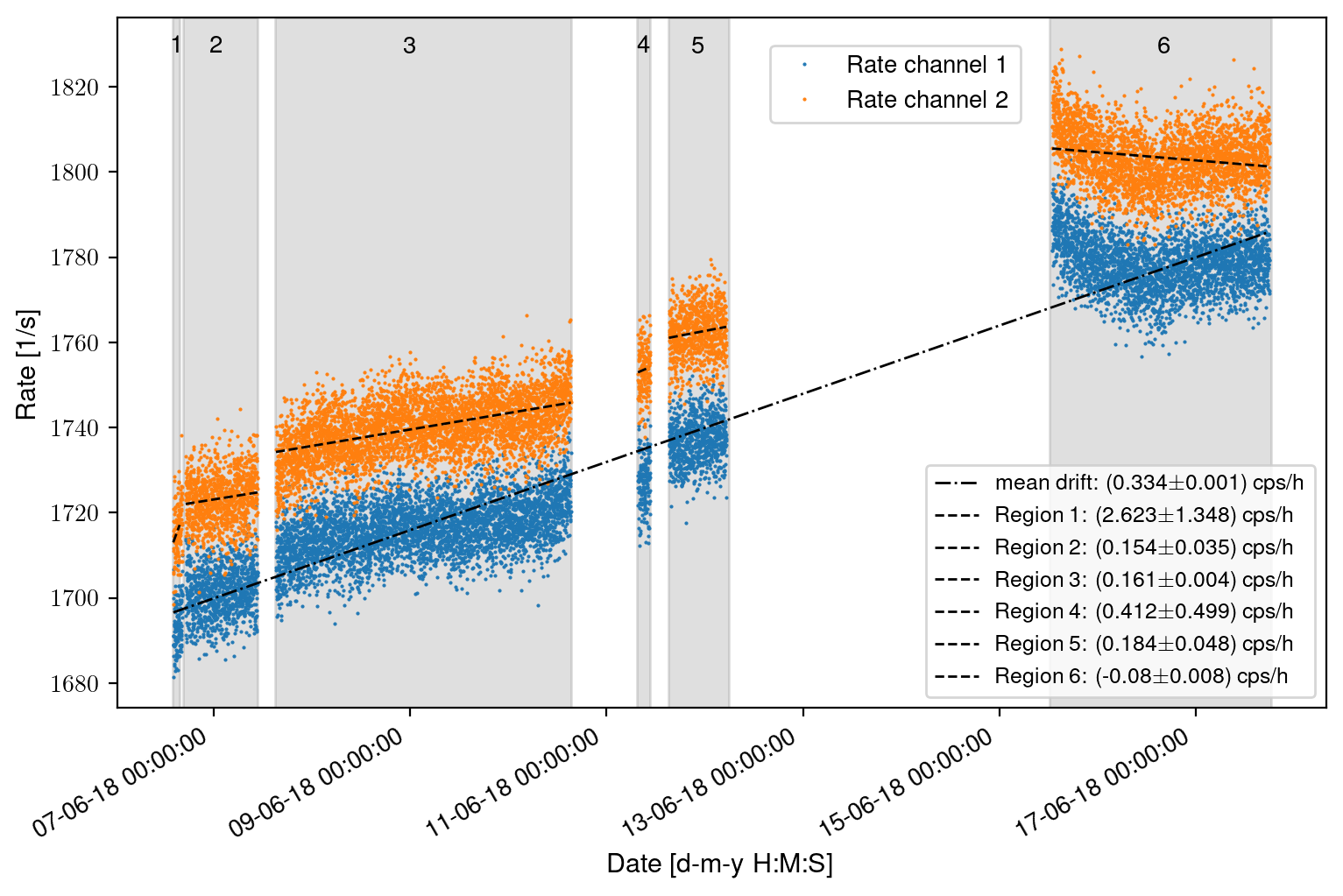}        
        \caption{$\upbeta$-electron rate trend summary of the first tritium campaign. The full available data from the stability measurements at the monitoring position for both channels is plotted. The count rates for channel 1 are approximately \SI{0.7}{\percent} smaller compared to channel 2 using the same energy threshold (here \SI{5.3}{\kilo\electronvolt}). Apart from that, the channels follow the same trend. The full linear fit reveals a mean relative increase of about \SI{0.02}{\percent} per hour while for the single regions this value is smaller than \SI{0.01}{\percent} per hour.}
        \label{Figure:Stability}
	\end{figure}
	
    During the two weeks of the first tritium campaign the FBM was mainly monitoring the flux in the CPS at position $x_{\text{FBM}} =$ \SI{65}{\milli\metre} (outer rim of the flux tube, see figure \ref{Figure:Fluxtube} and \ref{Figure:TritiumRateScan}). From time to time background measurements were taken slightly out of the beam at $x_{\text{FBM}} =$ \SI{80}{\milli\metre}. The full rate trend graphs are shown in figure \ref{Figure:Stability} for both detector channels including linear fits to the data. The entire monitoring time is separated into six time regions. There is a long term drift of approximately \SI{0.02}{\percent/\hour} determined from all regions, while for single regions the drift is generally smaller, especially for the longer regions \num{2}, \num{3}, \num{5}, and \num{6}, hence the reason for the larger long term drift must mainly originate from incidences which occur between the regions. Several investigations have been performed to find the source of this long-term drift, and there are hints that the detector response changes over time due to an increase in the noise level and degrading effects of the detector chip. Hence this drift is probably caused by the FBM and not by a change of the incoming electron flux. The latter assumption is supported by the results of the other monitoring systems which do not observe such a drift. However, this long-term drift is sufficiently small as the FBM is designed to monitor relative source fluctuations over short time intervals, such as seconds, minutes and at maximum a few hours. Within these time ranges the drift is within the required sensitivity of \SI{0.1}{\percent}. Therefore, despite the observed long term drift, the FBM shows a stability fulfilling its design goal.
    
    In the first tritium campaign of KATRIN the FBM was utilised to reduce systematic uncertainties in the tritium concentration $\epsilon_{\text{T}}$ measurement performed by the LARA system. In this campaign the tritium amount was limited to about \SI{1}{\percent} in deuterium, and consequentially statistical fluctuations in the determination of the concentration of the tritiated hydrogen isotopolouges were much stronger than it is the case for standard operation ($\epsilon_{\text{T}} >$ \num{0.95}). An average tritium concentration was determined over a long duration (\SI{\approx3}{\hour}) with LARA, and this average tritium concentration was fluctuated according to the higher statistics FBM data for short durations. Note that these are time-scales over which the drift of the FBM is negligible. This way the uncertainty of the short-term fluctuations measurement of the tritium concentration were reduced from about \SI{2}{\percent} down to about \SI{0.5}{\percent}. The cooperation of these two monitoring systems was crucial to reduce the tritium concentration systematic input for an upcoming \si{\kilo\electronvolt}-scale sterile neutrino analysis of the first tritium data.

	\section{Summary}
	\label{Section:Conclusion}
	
    The KATRIN experiment aims for a precise measurement of the electron antineutrino mass with a sensitivity of \SI{0.2}{\electronvolt} (\SI{90}{\percent} CL). One of the systematic uncertainties in this measurement arises from fluctuations of the column density of high luminosity tritium source. In order to reach the design goal of KATRIN, the latter must be measured on the per-mille level over time scales of a few minutes. Therefore the source is continuously monitored by several monitoring systems, one of which is the \textit{Forward Beam Monitor} (FBM). The FBM has the advantage of being capable of continuously monitoring variations of the electron flux and changes in the observed shape of the $\upbeta$-decay spectrum with high accuracy on short time scales.  
 
    A UHV compatible vacuum manipulator was commissioned. It is able to place a detector board directly into the beta-electron flux originating from the tritium source. Although the mounting position of the apparatus demands a movement mechanism with a working stroke of \SI{1.8}{\metre} the FBM is able to reach any position within the electron flux cross-section with a precision of better than \SI{0.3}{\milli\metre} which can be determined with magnetic field measurements.

    The detector board at the tip of the FBM manipulator measures the electron flux with two silicon \pin diodes. The FBM reaches an energy resolution of about $\sigma_{\text{FWHM}} =$ \SI{2}{\kilo\electronvolt} at an energy threshold of \SI{5}{\kilo\electronvolt}. The readout electronics are optimised to register electron events at a rate of $\mathcal{O}$(\SI{e4}{cps}) and thus to measure relative changes in the electron flux with \SI{0.1}{\percent} precision in about \SI{100}{\second}.

    The entrance window (dead layer) of the \pin diodes has a large impact on the detector response when measuring electrons. It was found that the dead layer thickness of the \pin diodes used for the FBM range from \SIrange{300}{500}{\nano\metre}.

    After commissioning, the FBM was employed for several KATRIN measurement campaigns. The capabilities of the FBM detector were confirmed as well as the positioning accuracy of the manipulator. A small long term (days to weeks) drift of the rate was observed which correlates to a drift of the noise level of the electronics. On short time scales (hours) the FBM is stable to the per-mille level. With this the FBM is a monitoring device which reaches all its design goals.

    With its good performance the FBM data already played a key role in reducing the systematic uncertainties of the tritium concentration $\epsilon_{\text{T}}$ fluctuations during the first tritium campaign. This was achieved by combining it with the LARA data which featured a relative statistical uncertainty of only a few percent on time scales of minutes due to the low amount of source gas molecules. This will be important for an upcoming \si{\kilo\electronvolt}-scale sterile neutrino analysis of the first tritium data.

  \acknowledgments 
  We acknowledge the support of the Ministry of Education and Research BMBF (05A14PX3, 05A17PX3) and the Helmholtz Association.

\end{document}

%% file: author-list.tex
\affiliation[a]{Department of Physics, Faculty of Mathematics and Natural Sciences, University of Wuppertal, Gauss-Str. 20, D-42119 Wuppertal, Germany}
\affiliation[b]{Institute for Data Processing and Electronics~(IPE), Karlsruhe Institute of Technology~(KIT), Hermann-von-Helmholtz-Platz 1, 76344 Eggenstein-Leopoldshafen, Germany}
\affiliation[c]{Institute of Experimental Particle Physics~(ETP), Karlsruhe Institute of Technology~(KIT), Wolfgang-Gaede-Str. 1, 76131 Karlsruhe, Germany}

\affiliation[d]{Institut für Kernphysik, Westfälische Wilhelms Universität Münster, Wilhelm-Klemm-Straße 9, 48149 Münster, Germany}
\affiliation[e]{Institute for Nuclear Physics (IKP), Karlsruhe Institute of Technology (KIT), Hermann-von-Helmholtz-Platz 1, 76344 Eggenstein-Leopoldshafen, Germany}

\author[b]{A.~Beglarian}
\author[,a]{E.~Ellinger\footnote{Corresponding author.}}
\author[a]{N.~Hau\ss{}mann}
\author[a]{K.~Helbing}
\author[c,a]{S.~Hickford}
\author[a]{U.~Naumann}
\author[d]{H.-W.~Ortjohann}
\author[e]{M.~Steidl}
\author[c]{J.~Wolf}
\author[b]{and S.~Wüstling}

\collaboration{KATRIN collaboration}

\emailAdd{ellinger@uni-wuppertal.de}